\documentclass[12pt,preprint]{aastex}

\newcommand{\ltsima}{$\; \buildrel < \over \sim \;$}
\newcommand{\simlt}{\lower.5ex\hbox{\ltsima}} 
\newcommand{\gtsima}{$\; \buildrel > \over \sim \;$}
\newcommand{\simgt}{\lower.5ex\hbox{\gtsima}} 

\newcommand{\feka}{\mbox{Fe  K$\alpha$}}
\newcommand{\fekb}{\mbox{Fe  K$\beta$}}

\newcommand{\xmm}{{\emph{XMM-Newton}}}

\newcommand{\lum}{erg~s$^{-1}$}

\newcommand{\nh}{cm$^{-2}$}
\newcommand{\nhsym}{N_{\mbox{\scriptsize H}}}

\newcommand{\sorg}{NGC~1365}
\newcommand{\chandra}{{\emph{Chandra}}}

\newcommand{\errUD}[2]{\ensuremath{^{+#1}_{-#2}}}

\newcommand{\fexxv}{Fe\,\textsc{xxv}}
\newcommand{\fexxvi}{Fe\,\textsc{xxvi}}
\newcommand{\suzaku}{{\emph{Suzaku}}}

\newcommand{\logxi}{erg cm s$^{-1}$}

\received{}
\begin{document}

\title{ \sorg: A low column density state unveiling a  low ionization disk wind. }

\author{V. Braito\altaffilmark{1,2}, J.~N. Reeves\altaffilmark{3,4}, J. Gofford\altaffilmark{3,4}, E. Nardini\altaffilmark{3}, D. Porquet\altaffilmark{5}, G. Risaliti\altaffilmark{6,7}
}
\affil{
$^1$INAF - Osservatorio Astronomico di Brera, Via Bianchi 46 I-23807 Merate (LC), Italy; valentina.braito@brera.inaf.it \\
${^2}$ASDC--ASI, Via del Politecnico, 00133 Roma, Italy\\
$^{3}$Astrophysics Group, School of Physical and Geographical Sciences, Keele University, Keele, Staffordshire ST5 5BG, UK\\
$^{4}$Department of Physics, University of Maryland, Baltimore County, Baltimore, MD 21250, USA\\
$^{5}$Observatoire Astronomique de Strasbourg, CNRS, UMR 7550, 11 rue de l'Universit\'e, 67000 Strasbourg, France\\
$^{6}$INAF - Osservatorio Astrofisico di Arcetri, Largo E. Fermi 5, 50125 Firenze, Italy \\
$^{7}$Harvard - Smithsonian  Center for Astrophysics, 60 Garden Street, Cambridge, MA 02138, USA\\
} 

 \begin{abstract}

  We present the time-resolved spectral analysis of the \xmm\ data of \sorg\, collected during one  \xmm\ observation, which  caught this ``changing-look''  AGN 
 in a high flux state characterized also by a low column density  ($\nhsym \sim 10^{22}$ \nh) of the  X-ray absorber.   During this observation  the low energy photoelectric cut-off is at about $\sim 1$ keV and  the primary continuum can be  investigated with the \xmm-RGS data, which show strong spectral variability that can be explained as a variable low $\nhsym$, which decreased from $\nhsym \sim10^{23}$ \nh\ to $10^{22}$ \nh\ in a 100 ks time-scale.
  The   spectral analysis of the last segment of the observation revealed the presence of  several absorption features that can  be associated with an  ionized (log $\xi \sim 2$ \logxi)  outflowing wind ($v_{\mathrm{out}} \sim 2000$ km s$^{-1}$). 
 We detected for the first time a possible  P-Cygni profile  of the Mg\,\textsc{xii} Ly$\alpha$ line associated with this  mildly ionized absorber  indicative of a wide angle outflowing wind. We suggest that this wind is  a low ionization zone of the highly ionized  wind present in \sorg,  which is responsible for the iron K absorption lines and is located within the variable X-ray absorber.
 At the end of the observation, we detected  a  strong absorption line at $E\sim 0.76$ keV   most likely associated with a lower ionization zone of the absorber (log $\xi \sim 0.2$ \logxi, $\nhsym \sim 10^{22}$ \nh), which   suggests that the variable absorber in NGC 1365 could be a low ionization zone of the disk wind.  
\end{abstract}

\section{Introduction} 

In the last decade several X-ray observations of Seyfert galaxies unveiled a  significant  variability of the column density ($\nhsym$) or covering fraction  of the X-ray absorbers (\citealt{Risaliti2002}), which  highlighted the complexity of the circum-nuclear medium of Active Galactic Nuclei (AGN).  The  emerging picture is that  the X-ray absorber of the  widely accepted Unified Model of AGN (\citealt{Antonucci})  is not a single and homogenous structure   located at a pc-scale distance from the central source,  but most likely multiple absorbers   co-exist. These  absorbers could be  located at different scales
at a pc-scale distance from the central nucleus like the putative torus as well as much closer in (i. e. at few tens of gravitational radii)  and  they  could be in part inhomogeneous (\citealt{Risaliti2002,Elvis04}). The advent of the modern X-ray observatory also unveiled  that photo-ionized or warm absorbers are common as they are observed  in at least 50\% of the unobscured AGN (\citealt{Reynolds1997,Porquet2004,Crenshaw2003,Blustin2005}). The warm absorbers give rise to several narrow absorption features (from several elements over a wide range of ionisation parameters), which when observed  at  high spectral resolution   are generally blue-shifted by a few hundred km s$^{-1}$ up to few thousand km s$^{-1}$, implying the presence of outflowing winds.  Recently, systematic studies of the X-ray spectra of bright nearby AGN unveiled  that  blue-shifted  absorption features due to  \fexxv\  and  \fexxvi\  are present in at least 40\% of the  radio-quiet AGN (\citealt{Tombesi2010a,Tombesi2011,Tombesi2012,Patrick2012,Gofford2013}) and also in a sample of local Broad Line Radio Galaxies (\citealt{Tombesi2010b,Tombesi2014}).  In comparison to the soft warm absorbers these winds are characterized  by high column densities ($\nhsym > 10^{23} $\nh), high ionization and high velocities ($v>0.1c$); thus their outflow rate can be huge (several $M_\odot$/yr) and  they can play a key role in the co-evolution of the  massive black hole and the host galaxy.

\sorg\ is a nearby ($z=0.0055$) Seyfert 1.8 galaxy,  and it has been the target of several X-ray  monitoring campaigns as it is the prototype of  the so called ``Changing-look'' AGN (\citealt{Matt2003}), where the column density of the X-ray absorber changes from Compton-thin (transmission dominated state; $\nhsym <10^{24}$ \nh) to Compton-thick (reflection dominated state; $\nhsym >10^{24}$ \nh). Furthermore, \sorg\ is also well known for the extremely rapid variability of its X-ray absorber;  the monitoring campaigns and the deep X-ray observations performed with \chandra, \xmm\ and \suzaku\   measured  a variability of the  column density (ranging from few $\times10^{23}$ \nh\ to $10^{24}$ \nh) and covering factor of the X-ray absorber  (from 10\% to 90\%) on the timescale of less than a  day-weeks (\citealt{Risaliti05Clook,Risaliti07,Risaliti09,Maiolino10,Brenneman13}). The emerging scenario for \sorg\ is that we are viewing the central X-ray source through an absorber made of clouds  (with a column density of few $\times10^{23}$ \nh\ up to $10^{24}$ \nh) located closer in than the pc-scale torus  and most likely  at a distance from the central black hole of the order of $10^{16}$ cm (\citealt{Risaliti07}), which is consistent with the distance of the  Broad Line Region (BLR) clouds.  The derived density of these clouds is also of the same order of the density of the BLR clouds (i.e. $n\sim 10^{10}$ cm$^{-3}$). Another important characteristic  of \sorg\ is  that  when it is  observed in  the Compton-thin state, its X-ray spectrum  shows  the persistent presence of strong  absorption lines due to  \fexxv\  (both the He-$\alpha$ and He-$\beta$) and \fexxvi\ (both the Ly-$\alpha$ and Ly-$\beta$ components); these lines are blue shifted  by $v \sim 3000$ km s$^{-1}$  (\citealt{Risaliti05,Brenneman13}). The intensity  and velocity shift of these lines  imply the presence of  a highly ionised (log\,$\xi \sim 3.5$\footnote{The ionization parameter is defined as $\xi=L_\mathrm{ion}/nR^2$, where $R$ is the distance of the ionizing source from the absorber, $n$ is the electron density and $L_{\mathrm{ion}}$  is the ionising luminosity in the 1-1000 Rydberg range.} \logxi) outflowing wind with column density of a few $10^{23}$ \nh\  and a possible  origin in an accretion disk wind. Recently  it has been proposed  that the long term variability of the X-ray absorber  could be explained as   a variable and  clumpy ionized wind that is responding to the changes in the accretion rate of \sorg\ (\citealt{Connolly2014}).\\

Despite the strong variability of the X-ray absorber and the possible variability of the primary power-law flux (\citealt{Brenneman13}), the  soft X-ray emission,  which is normally absorbed below 2 keV,   did not show any variability in either the spectral shape or in the intensity of the several emission lines   (\citealt{Guainazzi09}). The analysis of the 0.5 Ms high-resolution spectrum, obtained combining all the \xmm\ reflection grating spectrometer (RGS; \citealt{RGS_REF}) data (from the 2004 and 2007 observations),  showed that the soft X-ray emission is dominated by emission lines as observed in several Seyfert 2 galaxies (\citealt{Guainazzi07}).  However, the emission is not purely dominated by the AGN  components  but requires the presence of a significant contribution  from  a collisionally ionised plasma.  Several emission lines from  He- and H-like ions (from C to Si) as well as the L-shell transitions from Fe\,\textsc{xvii} were detected; this emission line component could be thus described as an  hybrid (collisionally + photoionized) gas.   The collisionally ionized gas was modeled with  two thermal components with temperatures of about $kT \sim 300$ eV and $ kT \sim 640$ eV (\citealt{Guainazzi09}), possibly associated with the  strong nuclear starburst activity as suggested  by the diffuse emission detected with \chandra\ (\citealt{Wang09}).  \\
 
Last year \sorg\ was  observed four  more times as part of a \xmm\ and NuSTAR (Nuclear Spectroscopic Telescope Array; \citealt{NUSTAR}) monitoring program; the observations were performed simultaneously to provide the optimal broadband X-ray spectra  to investigate the overall X-ray emission and spectral variability.  The results of the broadband analysis of the first observation (performed  on July 25-27, 2012) and the determination of the possible spin parameters of the central black hole were recently published in   Risaliti et al. (2013) and  \citet {Walton1365}.   During this  monitoring program  \sorg\ showed a new and even more dramatic spectral variability, which for the first time affected also the soft X-ray emission and which we consider in more detail here. While the first of these  observations caught the source in its standard Compton-thin state ($\nhsym\sim 3\times10^{23}$ \nh)  during the following observations (performed in December 2012, January 2013 and February 2013)  the primary power-law flux increased by a factor of 3 and more importantly  during the third observation the $\nhsym$ dropped to $\sim 10^{22}$ \nh (Rivers et al. in prep; \citealt{Walton1365}). The overall X-ray emission is now reminiscent of a weakly obscured AGN with the low energy photoelectric cut-off   at about $1$ keV.\\

 We thus examined each of the \xmm\ RGS  spectra  to investigate the possible signatures of this low column density absorber, our goal is to investigate   its ionization state   and infer its  location with respect to the variable high column density absorber. In this paper we will focus on the third observation (i.e. the one performed in January 2013), where,  as we will show,  \sorg\ was not only brighter but also showed for the first time a remarkable spectral variability in the RGS energy band ($0.3-1.8$ keV). In order to better understand the variable component we compared this observation with the historical RGS data and with our recently obtained  \chandra\ High-Energy Transmission Grating (HETG; \citealt{Markert94}) spectrum ($\sim 200$ ks exposure time)  of \sorg.  In all these past observations the AGN is fully absorbed below 2 keV and thus these spectra provide the template for the non variable  and distant emission line component.  The paper is structured as follows: in \S 2  we describe the \xmm-RGS and \chandra\  observations and data reduction; in \S 3 we present the RGS  spectral variability and the modeling of the least obscured segment of the observation. The discussion on the physical interpretation of this absorber is presented in \S 4.\\

\section{Observations and data reduction}

\subsection {\xmm}
Four  130 ks long \xmm\ observations of  \sorg\ were performed on July 2012, December 2012, January 2013 and February 2013. The EPIC-pn and MOS instruments were operating in full frame mode and with the medium filter applied. The results on the variability of the X-ray absorber during  these new \xmm-EPIC observations  will be presented in a forthcoming paper (Rivers et al. in prep.). 
The \xmm\ data
have been processed and cleaned using the Science Analysis Software   (SAS ver 12.0.1)  and analyzed
using standard software packages (FTOOLS ver. 6.13, XSPEC ver. 12.7; \citealt{xspecref}).   
The \xmm-pn data were filtered for high background time intervals which   yields  net exposure time
  of $\sim 93$ ksec.  The EPIC-pn  source and background  spectra  were extracted  using a
circular regions with a   radius of $35''$ and of $30''$, respectively.   Response matrices and ancillary response files at the source position  were created using the
SAS tasks \textit{arfgen} and \textit{rmfgen}.  The source spectrum was then  binned to have at
least 25 counts  in each energy bin. We did not use the EPIC-MOS data as  they are piled-up due to the high count rate of \sorg. \\

\begin{figure}
\begin{center}
\rotatebox{-90}{
\epsscale{0.6}
\plotone{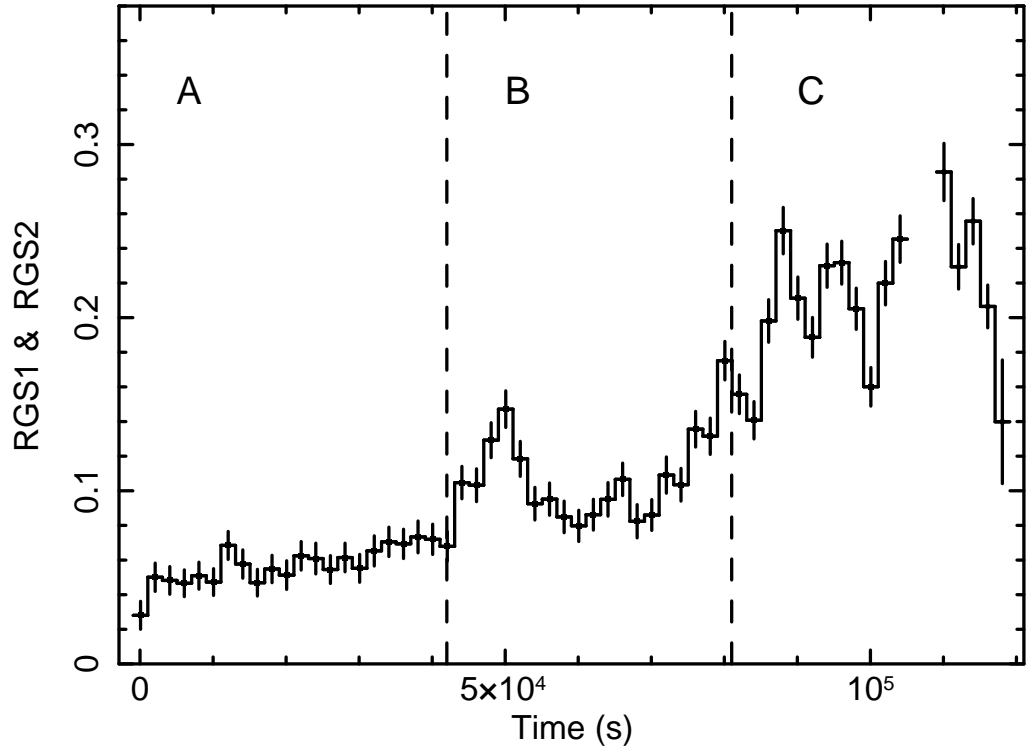}}
 \figcaption[curve_rgs_obs3paper.eps]{
RGS net light curve binned over 2 ks showing the increase of the soft X-ray flux during the third observation. The vertical dashed lines indicate the intervals that were chosen to investigate the spectral variability in the soft X-ray band. 
 \label{fig:rgs_curve}}
\end{center}
\end{figure}

For the scientific analysis   of this paper we  concentrated on the RGS  data,  which have the highest spectral resolution in the soft X-ray band  (0.35--1.8 keV)  with respect to the EPIC-CCD resolution. The RGS data have been reduced using the  standard SAS task {\it rgsproc} and the most recent calibration  files. High background time  intervals have been filtered out applying a threshold of  0.2 counts/s on  the background lightcurve  extracted  from a background region on the  CCD-9, which is the one more affected by  background flares and the closest to the optical axis.
We then extracted, for each observation, the background-corrected light curves   using the SAS task {\it rgslccorr} for the total  RGS band  adopting a binsize of 1 ks.
 We inspected both the light curves obtained for each RGS (considering only the first order data) as well as for the whole RGS (i.e. combining both the RGS1 and RGS2) and we  found a good agreement between the  RGS1  and RGS2 light curves. The inspection of these light curves shows strong variability in the RGS band only in the third observations. During this observation the two RGS collected a total of $\sim 12950$  net counts (0.35--1.8 keV).  In Fig.~\ref{fig:rgs_curve} we show the combined RGS light curve for this observation\footnote{We note that the gap present in the light curve is due to the splitting of the observation into two parts.}.   \\
 
 To assess if the observed variability  is caused by  fast variability of  the  low column density X-ray absorber,   we  divided this XMM-Newton observation into 3 intervals  of about 40 ks each (vertical dashed lines in Fig.\ref{fig:rgs_curve}). The intervals were chosen to have enough count statistics for a meaningful spectral analysis  and according to the increasing flux trend seen in the light curve. The first interval  shows only a weak increase of the RGS net counts, the second interval includes the first  ``flare''  and steady rise, while the last one is when stronger variability is present.  For each of these intervals we extracted the  RGS1 and RGS2   spectra  using the  standard SAS task {\it rgsproc}.  The two RGS collected a total of $\sim 2300$, $\sim 3950$ and $\sim 6700$ net counts  in the first, second and third interval, respectively.   The spectra were then binned  at half  the FWHM resolution of the instrument ($\Delta \lambda\sim 0.05\AA$)  and the C-statistic was employed in all the fits.

\subsection {\chandra}
\chandra\  observed \sorg\   twice in 2012, on April 09 and April 12, for a total of $\sim 200$ ks (OBSIDs: 13920 and 13921; see Table~\ref{tab:log_observ}). The observations were made  with the HETG instruments   at the focal plane of the Advanced CCD Imaging Spectrometer (ACIS-S; \citealt{Garmire2003}). The HETG consists of two gratings assemblies, the High-Energy Grating (HEG; 0.7--10.0 keV) and a Medium-Energy Grating (MEG; 0.4--8 keV). The HETG data were  reprocessed with the Chandra Interactive Analysis of Observations software package  (CIAO version 4.4\footnote{http://cxc.harvard.edu/ciao}; \citealt{Ciaoref}) and   CALDB version 4.4.9.  For each observation, events were extracted for each arm ($-1$ and $+1$) for the first order data of each of the HETG grating (HEG and MEG). 
Spectral redistribution matrices ({\it rmf} files) were made with the  CIAO tool {\it mkgrmf} for each arm ($-1$ and
$+1$); telescope effective area files were made with the  CIAO script {\it fullgarf} which drives the  CIAO tool {\it mkgarf}. Spectra were extracted from these events combining the $-1$ and $+1$ orders (using response files combined with appropriate weighting);  we did not subtract the background  as it is negligible. 
For the extraction we used a narrower extraction strip than the default (20 arcsec instead of 35), the reason for this choice is  to extend the HEG data above 8 keV and also to exclude part of the extended emission  from the starburst.  After the inspection that the spectra extracted from the two OBSIDs are consistent, we summed   them to create high quality summed first-order HEG and MEG spectra for fitting, again we created the   response files combining them with the appropriate weighting.   \sorg\ was in a Compton-thick state for the whole duration of the \chandra\ observation. For the scientific analysis   of this paper we  concentrated on the MEG  data, which partly overlap with the RGS spectra;  we note that despite the fact that  the source was in a Compton-thick state the MEG collected a total of $\sim 1700$  and $\sim 1100$ net counts in the 0.6--7. keV  and 0.6--1.8 keV energy range, respectively. For the spectral fit we then binned the  MEG to   2048   channels (half of  the FWHM of the spectral resolution), corresponding to  $\Delta \lambda \sim 0.01\AA$    and, as  for the RGS spectral fits, we employed the C-statistic.\\ Throughout the paper  we adopted  H$_0=71$ km s$^{-1}$ Mpc$^{-1}$, $\Omega_{\Lambda}$=0.73, and $\Omega_m$=0.27 \citep{Spergel2003}. 
All the  fit parameters are quoted in the rest frame of \sorg\ and errors are at the 90\% confidence level   for one interesting parameter, unless otherwise stated.

\section{Spectral Fitting}
\subsection{Variability of the Soft X-ray continuum}
\begin{figure}
\begin{center}
\rotatebox{-90}{
\epsscale{0.6}
\plotone{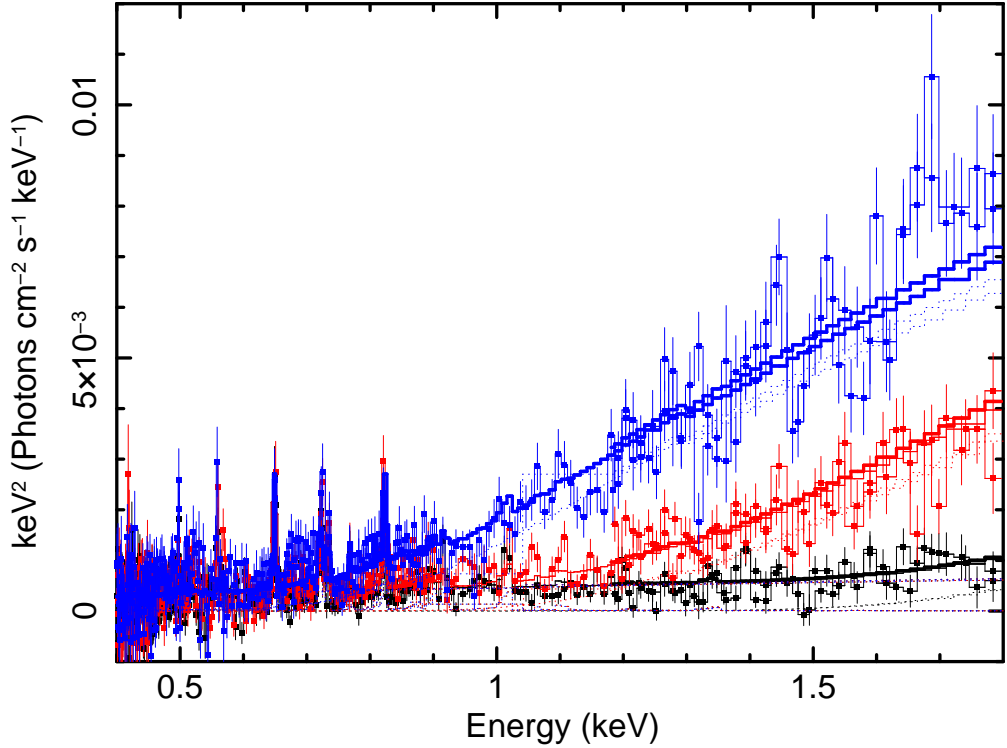}}
 \figcaption[slices]{
Fluxed RGS1 and RGS2 spectra obtained dividing the third observation into three intervals (black data points first 40 ks, red second interval, blue last 40 ks). The underlying continuum model  (solid line) is composed of a primary absorbed power-law component (dotted line), a scattered component and a collisionally ionized plasma (dashed line).  A clear decrease of the $\nhsym$ is  evident; while below $\sim 0.8 $ keV the spectra are similar they clear diverge above 1 keV due to the appearance of the primary emission. In the  spectrum collected during the last 40 ks several emission/absorption features are present. In particular in the $1.1-1.7$ keV the spectrum shows a clear imprint of a possible ionized absorber.
 \label{fig:slices}}
\end{center}
\end{figure}

As \sorg\ is known to have a highly variable X-ray absorber, we first tested if the variability shown in the RGS light curve could be explained as a variable low column density absorber. 
Thus we fitted the  three  RGS1 and RGS2 spectra with a simple baseline continuum composed  of:  a primary absorbed and redshifted power-law component,    a soft and unabsorbed power-law component (with the same photon index $\Gamma$) and a  thermal  emission model  ({\sc mekal} model in {\textsc{xspec}},  \citealt{Mewe85}). The latter component    represents the emission  due to a collisionally ionized plasma, which    from the previous \chandra\ and RGS analysis (\citealt{Wang09,Guainazzi09}) is expected to contribute to the soft X-ray emission from \sorg. We included also a  constant multiplicative offset  between the RGS1 and RGS2, which is found to be within $\pm 4$\% of 1.0.
In all the fits we included a Galactic column density of $\nhsym=1.34\times 10^{20}$\nh\  (\citealt{nhref}), adopting the   cross-sections and abundances of \citet{Verner} and  \citet{Wilms2000}. 
For all the intervals we tied  the photon index of the power-law component, while  we allowed  the  column density of the intrinsic absorber to vary. 
Although this fit is statistically unacceptable ($C/d.o.f.=3838.1/2818$) it provides a first order representation of the spectral variability. The resulting  fluxed RGS spectra  are shown in   Fig.~\ref{fig:slices}, where it is clear that the main driver of  the increase  of the soft X-ray flux is a decrease of the $\nhsym$ of the intrinsic absorber. When parametrized by this  single neutral absorber   and keeping tied the normalization of the power-law components,  its column density  changes from $\nhsym= 8.3\errUD{2.4}{1.2}\times 10^{22}$ \nh\ to $\nhsym=2.6\errUD{0.2}{0.2}\times 10^{22}$ \nh\ and  $\nhsym=0.9\errUD{0.1}{0.1}\times 10^{22}$ \nh\ for  the first, the second and the  last interval, respectively.  The photon index is found to be $\Gamma=1.52\pm0.14$ (over the RGS band) and the temperature of the thermal component is $ kT =0.34\pm 0.02$ keV, in good agreement with the value reported in \citet{Guainazzi09}.  We note that most of the remaining residuals  are due to emission and absorption lines, however statistically the fit does not require the presence of a second thermal component  as found in the previous work. We also tested   if the  difference in the spectral shape could be explained without a change of the column density of the neutral absorber, to this end we adopted the same baseline model but we allowed to vary only the photon index between the three intervals, the fit is worse by a $\Delta C= 1329.7$ and to account for the different curvature the photon index varies from  $\Gamma \sim 2.5$ to an unphysical value of $\sim -1.5$. We also allowed the normalization of the primary power-law component to vary, in this latter test the fit-statistic is better $\Delta C=68.2$ with respect to the variable absorber and the photon index varies from $\sim 1.4$ to $\sim 2.3$. However in this scenario the intensity of the primary continuum increases by a factor of 100 in the last interval, which is in disagreement with the variability seen in the broadband EPIC-pn spectra of this observation,  where the primary power-law flux varies only by a factor of 2 (see \S~ 3.5). \\

\subsection{The emission line spectrum}
As can be seen in Fig.~\ref{fig:slices}, in the third segment we  can observe the primary X-ray emission emerging above 0.8 keV and several absorption features are present in the $1.1-1.8$  keV energy range, which can be associated to transitions from highly ionized Ne and Mg and suggest the presence of a ionized absorber. Before attempting any investigation of this  absorber   we   proceeded to construct  a more complex baseline model that includes the narrow emission features present in the deep RGS spectra, obtained from summing all the observations collected from 2004 and 2007 with a total exposure time of about 500 ks for each RGS.  Thus we will be able to have a correct  parametrization of the absorption features  (i.e $EW$, $\sigma$ and velocity shift) from which we can infer the density, ionization state and location of the low $\nhsym$ absorber.
For the analysis we considered also the HETG spectrum, although the exposure time is  less than the combined RGS spectra the smaller extraction region provides a spectrum with less contribution from the diffuse  component and it extends to higher energies (e.g. $\sim 7 $ keV).
For these past observations as the $\nhsym$ is $>10^{23}$\nh\  we adopted as a primary continuum the scattered power-law component and the thermal emission component.  
Given that  in the third segment we see the primary continuum, we  fixed the photon-index of the  power-law component to the best fit value obtained from the broadband fit ($\Gamma=2.1$; see \S ~3.4) and we included successive narrow  Gaussian soft X-ray emission lines assuming their width  to be less than the instrumental resolution.   We fit simultaneously the HETG and RGS data and, as the extraction regions are different, we allowed for a different  cross normalization  between the RGS and HETG spectra. Furthermore,  as the smaller extraction region of the  HETG spectra includes  a lower contribution from the starburst  emission, we allowed for different normalizations of  thermal component and some of the strongest  Fe \textsc{xvii-xviii} emission lines in the 0.7--0.8 keV,  which are most  likely associated with the diffuse component. The  resulting emission lines component  is similar to the one presented in \citet{Guainazzi09} although for some lines we have a slightly different normalization that we ascribe  to the steeper photon index that was assumed in our analysis.  We note that   thanks to the MEG spectrum  above 0.6 keV we can deconvolve some of the emission lines into the putative AGN emission  and the  thermal emission components (Gofford et al. in prep).\\

\subsection{Absorption  lines  in the low column density state.}

\begin{figure}
\begin{center}
\epsscale{0.3}
\rotatebox{-90}{
\plotone{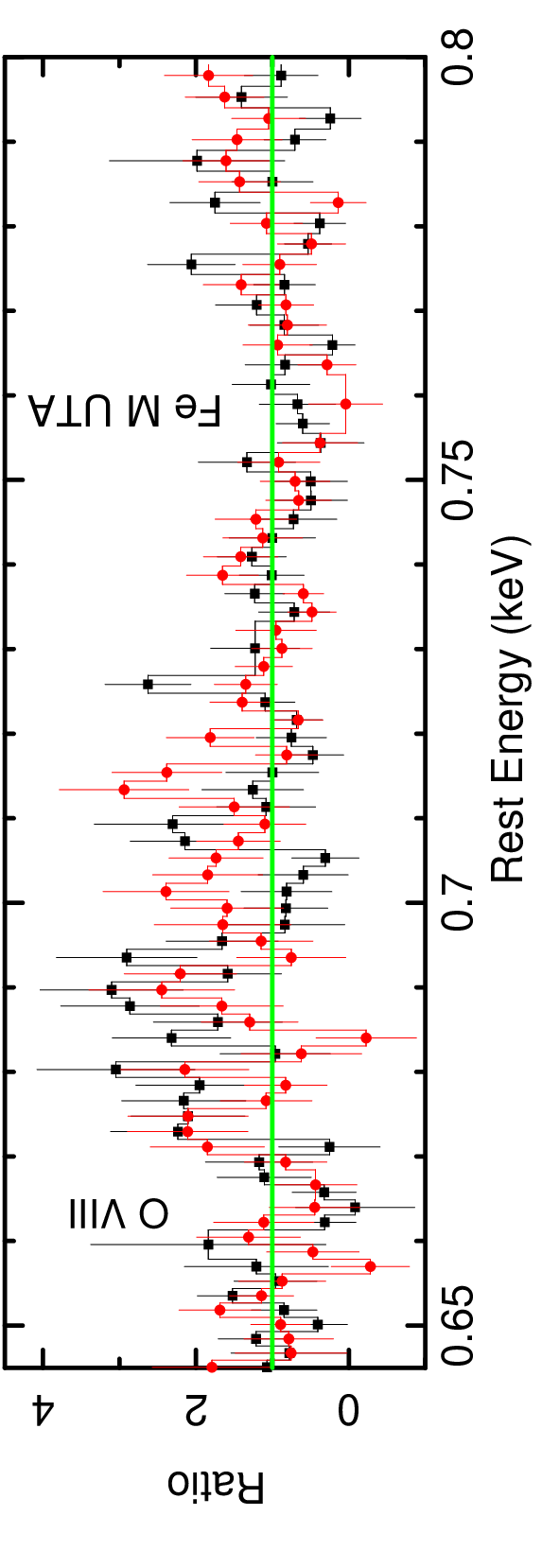}
\plotone{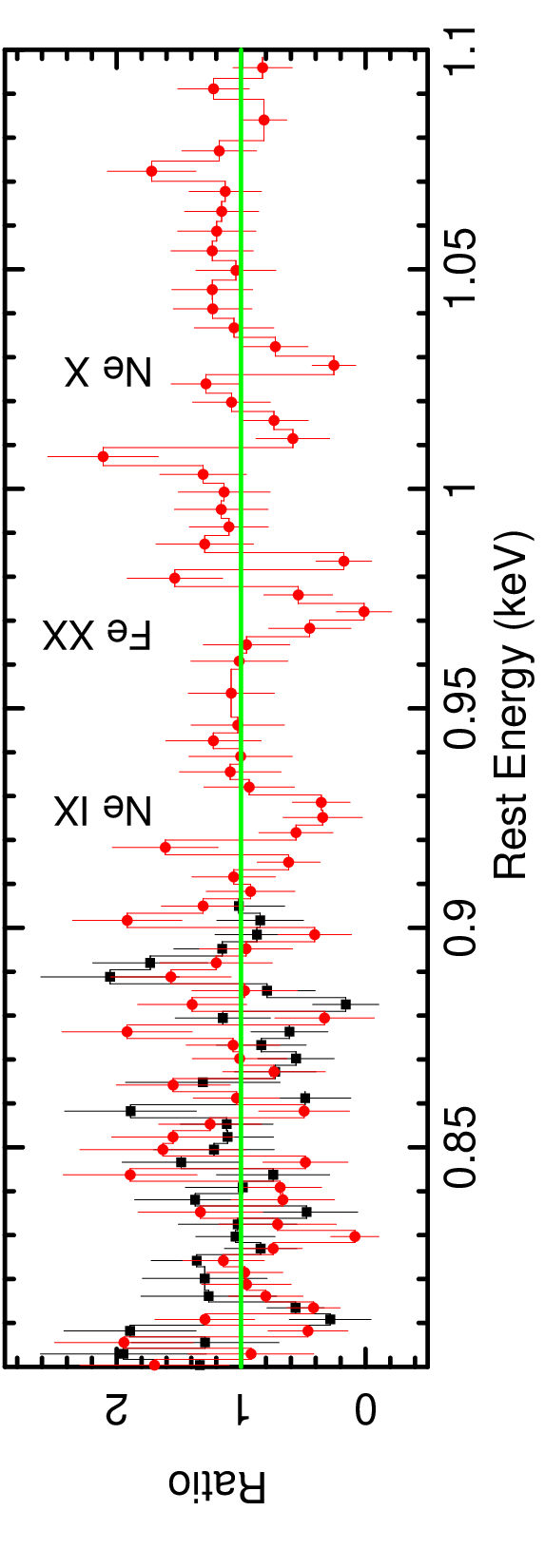}
\plotone{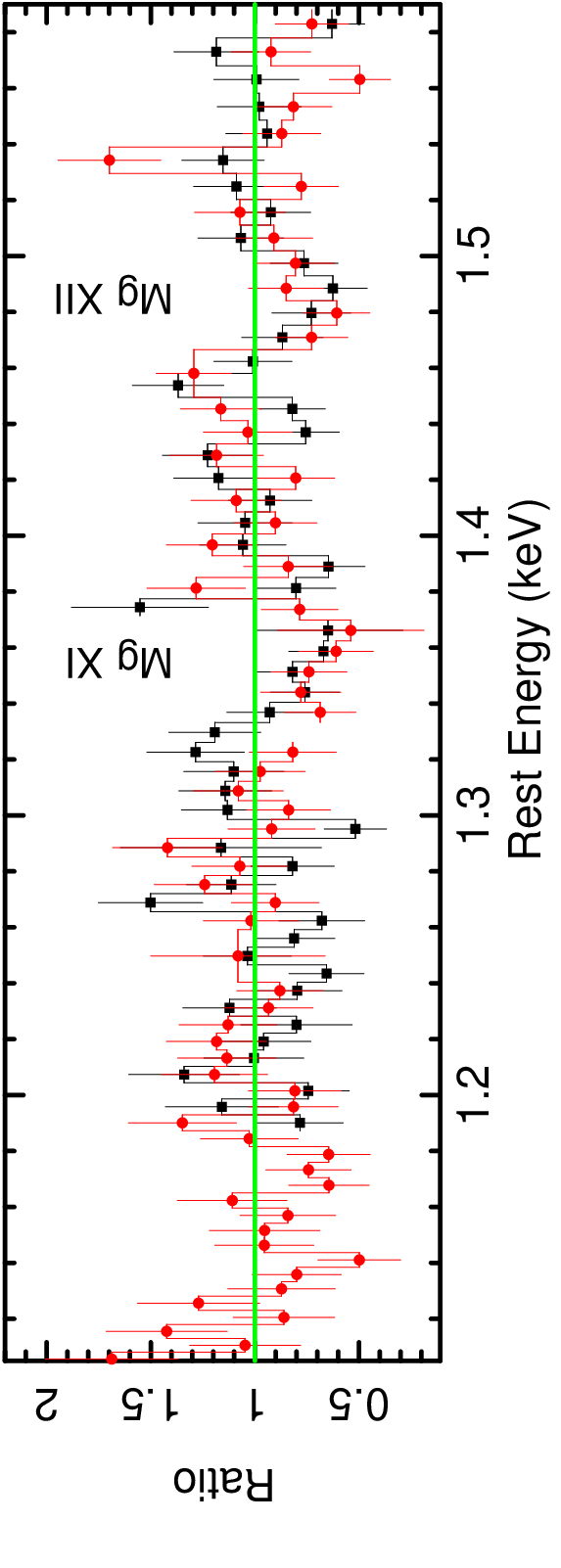}
}
 \figcaption[esiduals_rgs]{
Residuals in the RGS against the baseline model that includes the emission lines detected in the 2004-2007 RGS spectra. Several absorption lines can be clearly seen. The upper panel shows the Fe M-shell band, the middle panel shows the  absorption due the Ne\,\textsc{ix} and Ne\,\textsc{x};  the lower panels shows absorption Mg\,\textsc{xi} and Mg\,\textsc{xii}.
 \label{fig:residuals_rgs}}
\end{center}
\end{figure}

We thus applied this baseline model for the emission component  to the third interval and we  allowed  to vary  the intrinsic column density, the normalization of the power-law  component  and the thermal emission  component, while we kept fixed the emission line parameters.     As expected, we found that the thermal emission component does not vary, being produced in the extended starburst region.  Although this is a better representation of the  RGS spectra, it is still a poor fit  with an unacceptable statistic of $C/d.o.f.=1295.7/ 972$. Several  absorption features appear to be present  (see Fig.~\ref{fig:residuals_rgs}) at  $\sim 0.65$ keV (due to O\,\textsc{viii}), at  0.9-1.1 keV, likely due to Ne \,\textsc{ix} and  Fe\,\textsc{xx} and at  $\sim 0.7$ keV (Fe\,\textsc{xvii}-Fe\,\textsc{xviii}). Furthermore,  two broader absorption features are present at $\sim 1.3$ keV (Mg\,\textsc{xi}) and $\sim 1.5$ keV (Mg \,\textsc{xii}). We note that most of the residuals are blue-ward of the corresponding emission lines, and cannot be explained with an incorrect modeling of the emission line spectrum.  The only  emission line that requires a different normalization  with respect to the \chandra\ best fit is the Mg\,\textsc{xii} Ly$\alpha$.    We thus added several Gaussian absorption lines;  each  individual line was considered to  be statistically significant if the addition yielded an improvement of the fit statistic of  $C>9.2$ (corresponding to 99\% significance for 2 interesting parameters).  We then allowed their width to be free, but keeping the width tied to that of the  corresponding emission line. In particular the  width of the Mg \,\textsc{xii} absorption line  is tied to the Mg\,\textsc{xii} Ly$\alpha$ emission line, while for the Mg\,\textsc{xi} absorption line  we tied its widths to  the   Mg\,\textsc{xii} Ly$\alpha$,  as the  Mg\,\textsc{xi} is a triplet. For the Mg emission  and absorption  lines,   we found that  allowing their width to be free   we get  $\Delta C=7.9$ with respect to a fit with unresolved lines  and  $\sigma = 9\pm 3$ eV,  which corresponds to a velocity dispersion  of this absorbing gas of $\sigma = 1830\pm 610$ km s$^{-1}$ (or $FWHM\sim 4300$ km s$^{-1}$). We also allowed the widths of the Mg\,\textsc{xi}  and Mg\,\textsc{xii} to vary independently from the corresponding emission line, the fit does not improve ($\Delta C=1$) and we found  similar values of  $\sigma = 9_{-5}^{+8}$ eV and $\sigma = 9_{-5}^{+6}$ eV, for the Mg\,\textsc{xi}  and Mg\,\textsc{xii} absorption lines respectively.
\\

\begin{figure}
\begin{center}
\epsscale{0.45}
\rotatebox{-90}{
\plotone{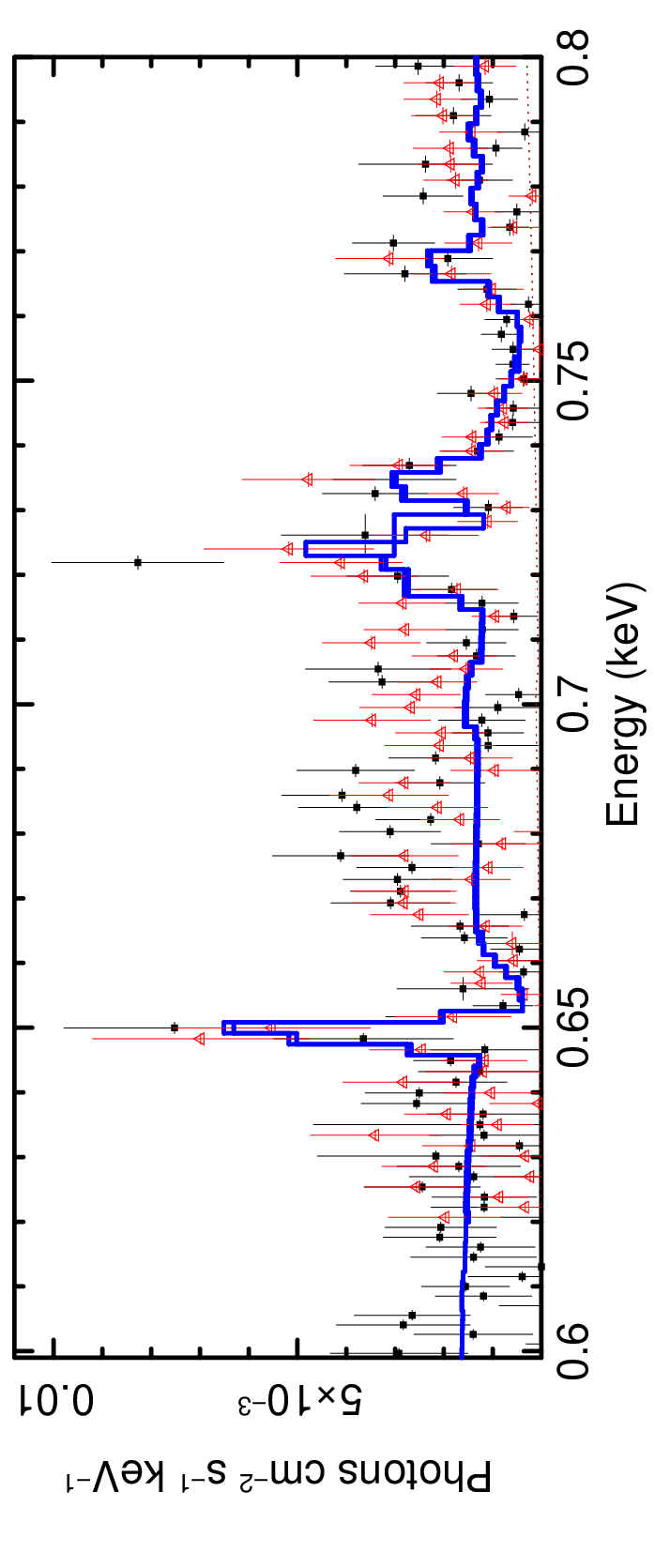}
\plotone{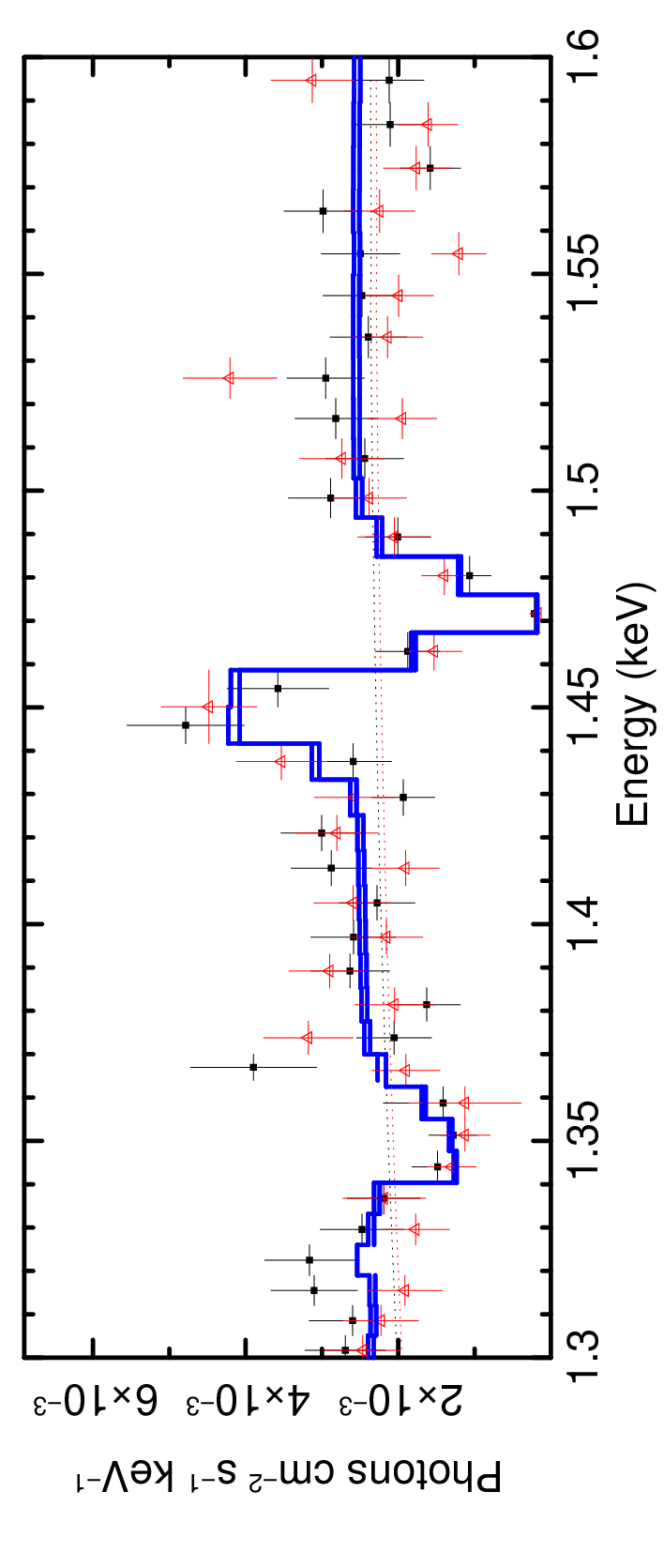}
}
 \figcaption[residuals_rgs]{
RGS unfolded spectrum showing  some of the absorption lines. The upper panel shows a zoom into the O\,\textsc{viii} and Fe M-shell region. The lower panel shows the 
 Mg\,\textsc{xi} and Mg\,\textsc{xii}  classical P-Cygni  profiles, composed of an emission and blue-shifted absorption  line. The profiles were fit with a double Gaussian model. 
 \label{fig:pcygni}}
\end{center}
\end{figure}

We detected 5 statistically significant absorption lines, in Table~\ref{tab:absline} we list their parameters together  with the most likely  identification, the corresponding laboratory energy and improvement of the fit, while in Table~\ref{tab:emlinergs} we  list the parameters of the strongest emission lines as detected in this segment of the third observation. Two   less significant  absorption  features are also present at the corresponding energies of the blue-shifted O \,\textsc{viii} Ly$\alpha$ and Ne\,\textsc{x} lines ($E \sim 657$ eV and $E\sim 1.029$ keV, respectively) but with a $\Delta C< 6$.  Although we note that within the errors  the   measured energies of the  strongest absorption lines are    consistent with the  rest frame energies, the energy centroids are  all blue-shifted with respect to systemic with an average shift of about $\sim 1000-2000$ km s$^{-1}$.  \\
The  profile  of the Mg\,\textsc{xii}  Ly$\alpha$ emission and absorption line, and to a less extent also the O\,\textsc{viii} profile,   is  reminiscent of a P-Cygni  profile (see  Fig.~\ref{fig:pcygni}) and thus   suggestive of a possible outflow.    
Albeit with the large uncertainties,   the  energy offset  between the Mg\,\textsc{xii}  Ly$\alpha$ emission  and  absorption line component   is   $\Delta E = 14\pm 7 $eV, where the emission line is slightly redshifted; if interpreted as    due to an outflowing   absorber  this energy shift would correspond to a velocity of this wind of  $v_{\mathrm{out}} \sim 2800$  km s$^{-1}$. Assuming that the emission and absorption lines  are indeed produced in the same outflowing gas the   similar intensity for the components of the   Mg\,\textsc{xii}  Ly$\alpha$ P-Cygni profile ($I\footnote{We note that later when we adopt an xstar grid for the photoionized absorber we will derive a lower intensity for this emission line.} _\mathrm{em}=6.5\errUD{1.6}{1.5} \times 10^{-5}$ ph cm$^{-2}$ s$^{-1}$and $I_\mathrm{abs}= -7.5\errUD{1.6}{1.5} \times 10^{-5}$ ph cm$^{-2}$ s$^{-1}$) is again  suggestive of a high covering factor of the outflowing  gas.  However the statistics of this short RGS spectrum prevent us from a more detailed investigation of the profiles as the line parameters are  
 subject to large  uncertainties. 
 
\subsection{Photoionization Modeling of the X-ray Absorber. } 
\subsubsection{The RGS spectra}

Taking into account the fact  that when \sorg\ is observed in a Compton-thin state  our line-of-sight intercepts an outflowing ionised absorber  (\citealt{Risaliti05}),  we attempted to model the soft X-ray spectrum replacing the absorption lines with a photoionized absorber.  We adopted a  model comprising a grid of photoionized absorbers generated by the \textsc{xstar} (version 2.2) photoionization code \citep{xstar}; this  absorber is placed in addition to the neutral  absorber. The grid covers a relatively wide range in ionization  (log $\xi$) and column density parameter space: $\nhsym=1\times 10^{20}-1\times 10^{24}$ \nh and log $\xi=0-6$ erg cm s$^{-1}$.   As some of the absorption lines appear to be resolved at the RGS resolution we adopted a grid with a turbulence velocity of 1000 km s $^{-1}$; this model provides an improvement of the fit of $\Delta C=22.3$  for 2 d.o.f. ($C=1268.3/969$), with respect to the model with no absorption lines. We then allowed   the absorber to be  outflowing; the fit improves for an additional   $\Delta C=13.1$  for 1 d.o.f. ($C=1255.2/968$). 
 However, this model cannot account for the absorption feature detected at $\sim 0.76$ keV, we thus included an additional absorption line ($\Delta C=20$, $C=1232.2/965$),  the line parameters are similar to one measured with the previous model.
The ionization is found to be log $\xi = 1.71\errUD{0.09}{0.07}$ erg cm s$^{-1}$,  the column density is $\nhsym = 1.9\errUD{0.8}{0.5} \times 10^{22}$ \nh\  and the outflow velocity is $v=2100\pm400$ km s $^{-1}$.  The additional absorption feature at $E=0.759\errUD{0.004}{0.003}$ keV could suggest the presence of a second lower ionization absorber as the line is at the corresponding energy of the Unresolved Transition Array (UTA)  due to $2p \rightarrow 3p$ transitions from  low ionization  M-shell iron (Fe less ionized than  Fe\,\textsc{xvii}). The presence of this feature generally implies the presence of an ionized absorber with log $\xi$ of order of $0-1$ \logxi.   We thus tested for a double ionized absorber and despite  the low exposure of these RGS spectra, we found  that the  $E\sim 0.76$ keV  absorption feature can be accounted for by a  low ionization  absorber.  The fit improves  by  $\Delta C=60.2$  ($C=1175.0/965$) with respect to the model with the $E\sim 0.76$ keV  Gaussian absorption line. For the higher ionization component  (zone 2 in Table~\ref{tab:ionabs}) we found a higher ionization parameter  (log $\xi = 2.1\errUD{0.1}{0.1}$ erg cm s$^{-1}$ and $\nhsym = 1.1\errUD{0.4}{0.3} \times 10^{22}$ \nh) and a slightly lower velocity ($v=1700\pm450$ km s $^{-1}$). For the low ionization zone  (zone 3 in Table~\ref{tab:ionabs}) we found  log $\xi = 0.2\errUD{0.1}{0.1}$ erg cm s$^{-1}$,  $ \nhsym = 1.1\errUD{0.1}{0.1} \times 10^{22}$ \nh;  its  velocity is unconstrained  ($v<2300$ km s $^{-1}$).  We note that  with this low ionization absorber  we do not require anymore the presence of the fully covering neutral absorber $\nhsym < 10^{21}$ \nh, suggesting that the variable soft  absorber always seen in \sorg\ could be a low ionization zone of the disk wind.

\begin{figure}
\begin{center}
\epsscale{0.6}
\rotatebox{-90}{
\plotone{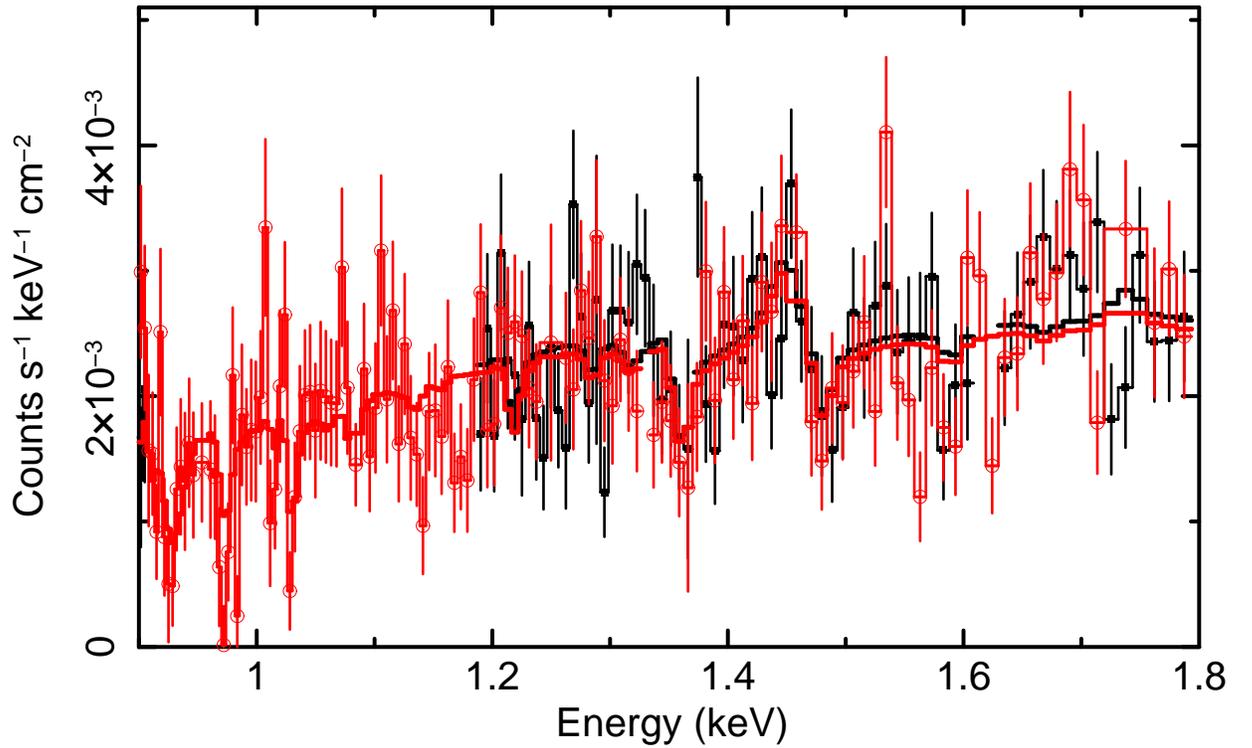}
}
 \figcaption[residuals_rgs]{
 Zoom into the 0.9--1.8 keV  RGS spectra (RGS1: black data filled  points, RGS2: red open data points), with the \textsc{XSTAR} best fit model.  The  model  for the X-ray absorber is composed of two ionized absorbers with a similar column density of $\nhsym \sim 10^{22}$cm$^{-2}$  (log $\xi \sim 2$ \logxi and log $\xi \sim 0.2$ \logxi; see Table~\ref{tab:ionabs}). 
 \label{fig:euf_rgs}}
\end{center}
\end{figure}

The best fit-parameters of this model are listed in Table~\ref{tab:ionabs}, while  the RGS spectrum with the best fit model of \sorg\ is  shown in Fig.~\ref{fig:euf_rgs}. It is worth noting the similarity of the outflow velocity of the  zone 2 mildly ionized absorber with the velocity generally measured for the highly ionised  iron K absorber (or zone 1 see below and Table~\ref{tab:ionabs};   $v_\mathrm{out}\sim 2000-4000$ km s$^{-1}$; \citealt{Risaliti05,Brenneman13}). \\
\subsubsection{The EPIC-pn spectra}
Given the parameters of this ionized absorber we expect that absorption features due to  He- and H-like  Mg, Si and S   should be present in the 1--3 keV energy range, we thus inspected the EPIC-pn spectrum extracted in the same  time interval.  Since we know from previous X-ray observations that the X-ray absorber is more complex than a fully covering absorber, we included in the model a neutral  partial covering absorber with no net outflow velocity. As for the RGS fit, we fitted   the baseline model with the line emission component   and we  allowed  to vary  the intrinsic column density of all the absorbers, the normalization of the power-law  component  and the thermal emission  component, while we kept fixed the emission line parameters; we also allowed to vary the photon index of the power-law component and the temperature of the thermal emission.  The model can be parametrized as:
\begin{equation}
 F(E) = {\rm phabs} \times [{\rm pow}_{\rm uncov} + {\rm Gauss} + {\rm \textsc{mekal}} + ({\rm zpcfabs} \times {\rm zphabs}\times {\rm pow}_{\rm cov})]
 \end{equation}
 
 where phabs is the Galactic absorption, ${\rm pow}_{\rm uncov}$ is the primary scattered power-law component,  ${\rm Gauss}$ are the soft X-ray emission lines, \textsc{mekal} is the thermal emission  component, $ {\rm zphabs}$ is a fully covering local absorber and  zpcfabs is a partial covering neutral absorber. \\ 

\begin{figure}
\begin{center}
\epsscale{0.7}
\rotatebox{-90}{
\plotone{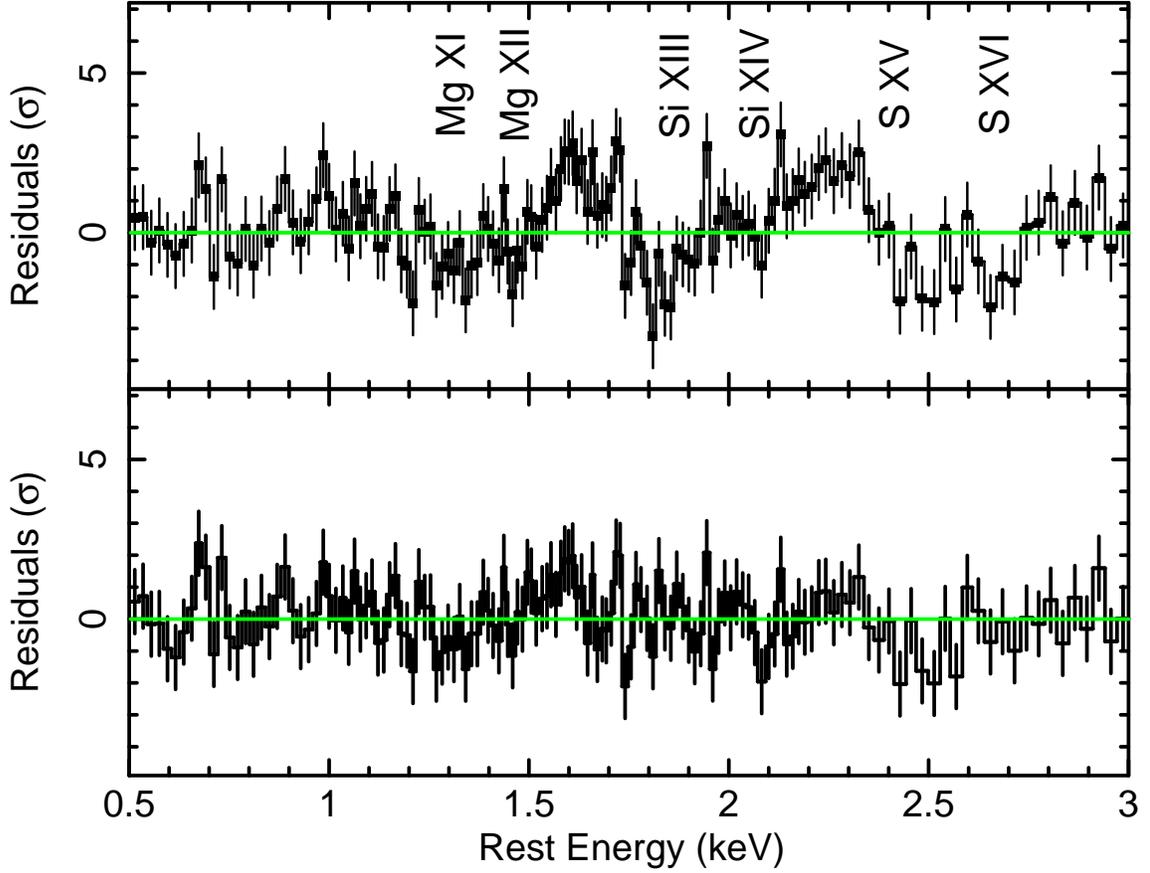}
}
 \figcaption[residuals_pn]{
Upper panel: Zoom into the 0.5--3 keV EPIC-pn  residuals (in $\sigma$) against the emission line model. The absorption lines detected in the RGS energy range are also clearly present in the EPIC-pn spectrum, furthermore some higher energy absorption features are present. The  strongest absorption features  are labelled. Lower panel: Zoom into the 0.5--3 keV EPIC-pn  residuals (in $\sigma$) against the best fit model which includes   three ionized absorbers. 
 \label{fig:residuals_pn}
}
\end{center}
\end{figure}

This model provides a poor fit to the 0.5--10 keV spectrum ($\chi^2/d.o.f.=1581.6/1264$),  and leaves residuals at the Fe K emission line region due to the \feka\ emission  line and the blue-shifted absorption lines. The fit is also poor   below 3 keV (see Fig.~\ref{fig:residuals_pn}),  confirming the prediction of the ionized absorber model  that was fitted to the RGS spectra, for the presence of absorption signatures due to  He- and H-like Mg,  Si and S.  In Table~\ref{tab:pnabslines} we report the strongest absorption lines detected in the EPIC-pn spectrum. Thus we added to the model a photoionized absorber, we also included the expected  narrow \feka\  emission line; the fit improves by  $\Delta \chi^2=80$  ($\chi^2/d.o.f.=1501.6/1258$) and   the ionization parameter is found to be similar to the value measured with the RGS  spectra log $\xi = 1.4\pm0.2$ \logxi. Although the fit improves, this photoionized absorber is not able to account for the absorption due to  highly ionized Fe, as the ionization  is too low. Similarly if we constrain the ionization to be greater than log $\xi >2$ \logxi\ the fit improves by a factor of $\Delta \chi^2=117.9$  (log $\xi = 3.8\pm0.1$ \logxi\ ; $\chi^2/d.o.f.=1383.7/1258$) with respect to the low ionization one, but we are not able to account for the absorption lines detected below 3 keV. \\

We thus tested a model with two photoionized absorbers, allowing  both their $\nhsym$ and $\xi$ to vary but assuming that they have the same outflow velocity. We also included in the model an additional broad Gaussian emission line at $\sim 6 $ keV to phenomenologically account for the relativistic Fe K$\alpha$ emission line (\citealt{Risaliti2013}) and the \fekb\ emission line. For the \fekb\ emission line we fixed the energy centroid to $E=7.06 $ keV and we tied its width to the width of the narrow \feka\ emission line  ($\sigma \sim 50$ eV) and its normalisation to be 13.5\% of the \feka\ emission line intensity  (\citealt{Palmeri}).

The model can be parametrized as:
\begin{equation}
 F(E) = {\rm phabs} \times [{\rm pow}_{\rm uncov} + {\rm Gauss}+  {\rm \textsc{mekal}} + ({\rm zpcfabs} \times {\rm zphabs}\times {\rm XSTAR}_{\rm Low}\times {\rm XSTAR}_{\rm Fe} \times {\rm pow}_{\rm cov})]
 \end{equation}
where now   ${\rm Gauss}$   includes both the soft X-ray emission lines and the Fe K$\alpha$ emission line complex, \textsc{mekal} is the thermal emission  component,  \textsc{XSTAR}$_{\rm Low}$ and \textsc{XSTAR}$_{\rm Fe}$ are  the low and high ionization  absorbers, respectively. As  with the previous model  
$ {\rm zphabs}$ is a fully covering absorber and  zpcfabs is a partial covering neutral absorber. \\

  This model  improves the fit by $\Delta \chi^2=44.5$  for 5 $d.o.f.$  with respect to a single ionized absorber  ($\chi^2/d.o.f.=1339.2/1253$).  The parameters of the higher ionization absorber are  similar to the historically reported values (log $\xi=3.7\pm{0.1}$ \logxi\ and $\nhsym =1.5\errUD{0.8}{0.5} \times10^{23}$\nh; \citealt{Risaliti05,Brenneman13}); the outflow velocity is  $v_{\mathrm{out}}=3500^{+1000}_{-700}$ km s$^{-1}$, while for the lower ionization zone  we found  $\nhsym= 8.4\errUD{2.6}{2.4}\times  10^{21}$ \nh\  and  log $\xi= 1.5\pm0.2$ \logxi.  The ionization parameter for the lower of the  two zone   model for the pn data is  intermediate  between the values of the two  RGS zones, suggesting that   it is trying to fit a mix of zone 2 and zone 3 as found with the RGS data.   We also allowed the two ionized absorbers to  have a different outflowing velocity and we found that for the low ionization absorber we do not require any net velocity. As the higher spectral resolution  of the RGS allows a more precise measurement of the outflow velocity of the low ionization component of the wind, we tested a model with the  velocity of the low ionization absorber fixed to the RGS value, while we allowed   the velocity of the higher ionization one to vary. Statistically the fit is equally good   ($\chi^2/d.o.f.=1330.4/1253$)  and the only parameter that varies with respect to the previous model is the outflow velocity of the high ionization absorber, which is now  $v_{\mathrm{out}}=3900^{+800}_{-800}$ km s$^{-1}$. We thus decided to keep for the following fits the  low $\xi$ velocity fixed to the RGS value.\\

Although  the fit is already statistically good there are still some residuals in the $1.5-3$ keV energy range possibly due to emission/absorption  lines from Si  and  S, which are not accounted for by this model; thus we  added several narrow Gaussian emission/absorption lines.  As for the RGS analysis an individual line was 
considered  to be statistically significant if the $\chi^2$ improved by a  $\Delta\chi^2>9.2$, corresponding to 99\% significance for 2 interesting parameters. The lines were assumed to be  unresolved at  the resolution of  the EPIC-pn. Two more Gaussian absorption lines are  required at $E \sim 2.5$\,keV ($\Delta\chi^2=23.3$) and at $E\sim 1.83$ keV ($\Delta\chi^2=49.9$), the former line could be identified with blue-shifted  He-like S ($E_\mathrm{lab}=2.461$ keV), while the latter is Si\,\textsc{xiii} ($E_\mathrm{lab}=1.83$ keV). An identification of the first absorption line with the He-like S would imply an outflow velocity of $\sim 3500$ km  s$^{-1}$.   Although we note that within the error  the   measured energy of this    absorption line is    consistent with the  rest frame energies,  the  outflow velocity is  similar to the outflow velocity of the higher ionization absorber but would require a lower ionization level (log $\xi \sim 2-2.5$ \logxi), which is intermediate between the ionizations of the two-zone pn model.  The presence of these two  absorption features suggests either the presence of a third component of the ionized absorber or  different abundances with respect to solar as assumed for the \textsc{xstar} grid.  We note that in the analysis of the \suzaku\ observation of \sorg\ Gofford et al. (2013) suggested the presence of a partial covering and ionized absorber with a similar ionization level (log $\xi\sim 2.4$ \logxi; see appendix D of \citealt{Gofford2013}).     We still require the presence of both a  fully and  a partial covering neutral absorber with  $\nhsym=(1\pm 0.1)\times10^{22}$ \nh\ and  $N_{\mathrm {H\;PC}}= 1.1\errUD{0.3}{0.3}\times 10^{23}$ \nh,  respectively and a covering fraction for the latter of $30\pm10$ \%.  The parameters of the   ionized absorber  are unaffected  by the addition of these Gaussian absorption lines  and consistent within the errors.   We still find that the ionization and  column density of the lower ionization zone  (log $\xi=1.3\pm{0.2}$ \logxi\ and $\nhsym =0.7\pm 0.2 \times10^{22}$ \nh) in the pn to be intermediate between the  RGS values.\\

We thus tested for a third component of the ionized absorber;  since the RGS analysis suggests that the partial covering absorber could be a  low ionization zone of the wind, we 
 replaced the neutral partial covering absorber with a partial covering ionized absorber with no net outflowing velocity. The fit improves with respect to the two ionized absorbers and one neutral partial covering absorber  ($\chi^2/d.o.f.=1292.2/1253$), while it is worse by a $\Delta \chi^2=34$ with respect to the phenomenological model with the Gaussian absorption lines. We found that the covering fraction of this third zone of the ionized absorber is  $94\pm 4$\%. Interestingly,  the column density  and ionization parameter of the lower ionization zone are now  log $\xi=0.6\pm0.3$ \logxi\ and $\nhsym =1.3\pm 0.3 \times10^{22}$ \nh, which  are in better agreement  with the parameters of the low ionization absorber   in the  RGS    (zone 3 in Table~\ref{tab:ionabs}); while the parameters of the mildly ionized absorber  are in good agreement  with the zone 2 absorber in the RGS (see  Table~\ref{tab:ionabs}). \\
 
 To summarize we have evidence for the presence of three ionized absorbers;  a highly ionized  and outflowing  zone (zone 1; log $\xi\sim3.8$ \logxi, $v_{\mathrm{out}}=3900^{+800}_{-1000}$ km s$^{-1}$), responsible for the iron K absorption features and seen only in the pn data. A mildly ionized absorber seen both in the RGS and pn data (zone 2; log $\xi\sim2.1$ \logxi) responsible for the Mg, Si and S absorption lines, which  is also outflowing ($v_{\mathrm{out}}=1700^{+400}_{-500}$ km s$^{-1}$ from the RGS data); then a low ionization absorber (zone 3; log$\xi<1$ \logxi) is responsible for the UTA, seen in the RGS,  and the overall spectral curvature of the pn spectrum.  Furthermore,  we found that the addition of a neutral partial covering absorber improves the fit by only $\Delta \chi^2=11$ for 2 $d.o.f.$ and thus is no longer strictly required. The main parameters of this best fit are reported in Table~\ref{tab:ionabs},  while the residuals are shown in Fig.~\ref{fig:residuals_pn} (lower panel). The only weak residuals left are around   $E\sim 0.65$ keV and $E\sim 1.6$ keV, which could be ascribed to a slightly  higher  normalizations of the   O\,\textsc{viii} Ly$\alpha$  and Mg\,\textsc{xi} Ly$\beta$ emission lines in the pn spectrum with respect to the RGS best-fit. We allowed the parameters of these two emission lines to vary and we found only a marginal improvement of the fit ($\Delta\chi^2=19$ for 4 $d.o.f.$).

\subsection{The broadband variability}

As shown by the RGS analysis the  main driver of the variability detected in the soft X-ray band can be explained with an uncovering of the primary X-ray source, which unveiled the presence of a  mildly ionized outflowing absorber  as well as the presence of emission-absorption lines with the classical P-Cygni profile.   During this observation   we measured with the RGS a decrease of the column density of the partial cover absorber  from $\nhsym\sim 10^{23}$ \nh\ to $\nhsym\sim 10^{22}$ \nh\  in a 130 ks time-scale. We thus investigated the variability of the X-ray absorber focusing on the low ionization one in the  broadband EPIC-pn spectra.  
We divided the observation adopting the same three intervals highlighted with the RGS ligth-curves. The first inspection of these spectra suggests a variability of the curvature that can be ascribed to the well established variability of the X-ray absorber, as well as a possible variability of the intensity of the primary continuum. \\

\begin{figure}
\begin{center}
\epsscale{0.7}
\rotatebox{-90}{
\plotone{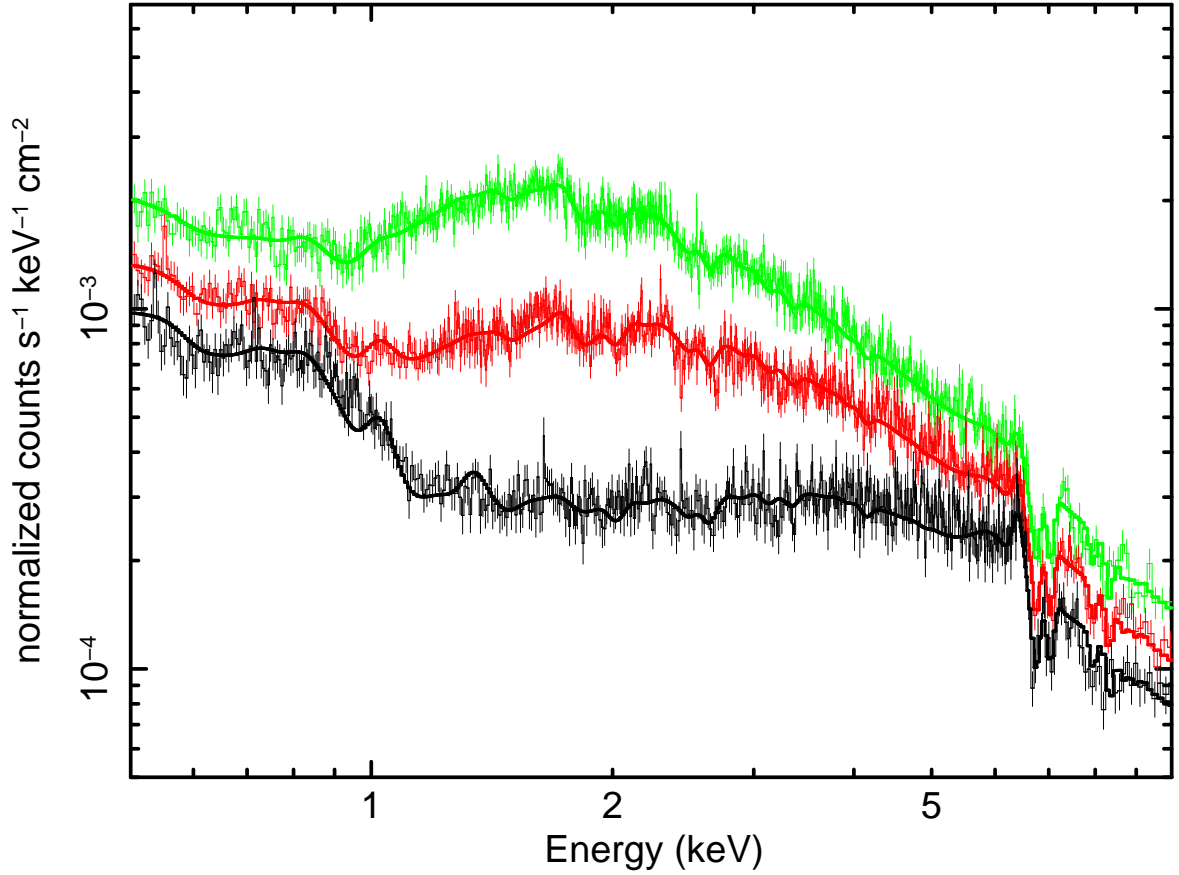}
}
 \figcaption[ev_pn]{EPIC-pn spectra  of the  three  intervals obtained dividing the  observation  with the same GTI adopted for the RGS analysis (black data points first 40 ks, red second interval, green last 40 ks).  The best-fit model  for the X-ray absorber is composed of: a fully covering   neutral  absorber, two ionized and outflowing absorbers and a ionized partial covering absorber. 
A decrease of the  amount of absorption  of the low ionization one is evident with a change in  $\nhsym$ from  $\sim  7 \times 10^{22}$ cm $^{-2}$  to $\nhsym$   $\sim  8 \times 10^{21}$ cm $^{-2}$. 
 \label{fig:euf_pn}
}
\end{center}
\end{figure}

  We fitted simultaneously the three   EPIC-pn spectra with the best-fit model found for the third segment with the 3-zones ionized absorber. As we see a clear variation of the spectral curvature both in the RGS and pn data we allowed to vary    the $\nhsym$ of the fully covering  neutral absorber and the column density of the soft X-ray absorber (zone 3), while we kept tied its covering fraction (best fit value of $87\pm 1$\%, see Table~6).  We note that the  covering factor is slightly lower than the best fit value found while fitting only the third interval;  this is most likely due to the  assumption that it is constant during the observation.
We also  allowed to vary  the normalization of the   primary  power-law component as the inspection of the spectra shows (see Fig.~\ref{fig:euf_pn}) a different flux  above 8 keV.  At first we kept tied the parameters of the highly   and mildly ionized absorbers (zone 1 and zone 2); with this model we can test if the mildly ionized  absorber could be always present  but it only  appears when the partial coverer absorber (zone 3) has a lower column density. Finally we allowed to vary the intensity  of the Mg\,\textsc{xii}  emission line, as    the intensity  measured with the RGS spectra is different with respect to the  measurement obtained with the \chandra\ data, and that there is an indication of variability of the emission lines in the RGS spectra (see below and Table 7).\\
This model already provides a good fit to  the three spectra ($\chi^2/d.o.f.=3909.3/3664$),  the inspection of the residuals shows that there are only   some weak residuals  possibly due to a change also of the  ionization of the low ionized absorber or covering factor. We  found evidence of  variability of the continuum (primary  and scattered  power-law components);  statistically if we constrain these intensities to be constant the fit is worse, in particular even  allowing for a variability  of  all the  three ionized absorbers ($\nhsym$ and $\xi$) we cannot account for the observed differences ($\Delta\chi^2$ is worse by  $ 860$).  We found a   variability of the  $\nhsym$ of the fully covering neutral absorber  (from $\nhsym=  (1.7\pm 0.2)\times 10^{22}$ \nh\  to  $\nhsym= (0.7 \pm 0.1)\times 10^{22}$ \nh); however the main driver of the  spectral variability of \sorg\ is  a change in the overall column density of the low ionization zone (zone 3); which during this observation drops from $\nhsym= (6.6 \pm 0.4)\times 10^{22}$ \nh\  to $\nhsym= (0.8 \pm 0.2)\times 10^{22}$ \nh. \\

In Table~\ref{tab:ionabs_pn} we report the best fit values of this model for the three segments. We note that when fitting the three intervals keeping tied the parameters of the  highly ionized wind,  the  outflow velocity appears to be  smaller than the value measured in the last part of the observation.  The short time scale variability of the highly ionized wind will be presented in Nardini et al. in prep.  however, we stress  that the main driver of the spectral  variability is a change in the column density of the    partial covering absorber (see Fig.~\ref{fig:euf_pn}). \\

\begin{figure}
\begin{center}
\epsscale{0.5}
\rotatebox{-90}{
\plotone{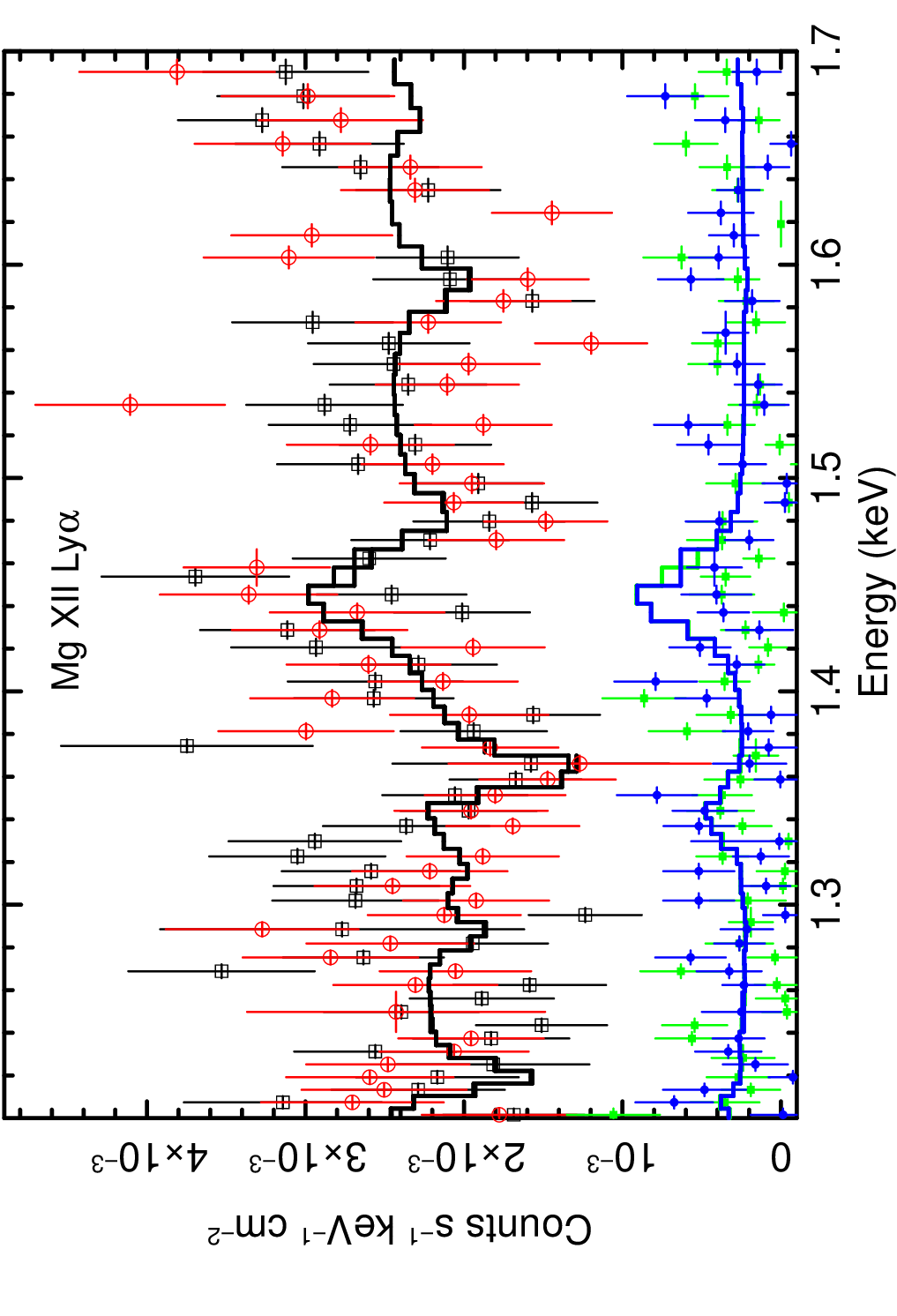}
\plotone{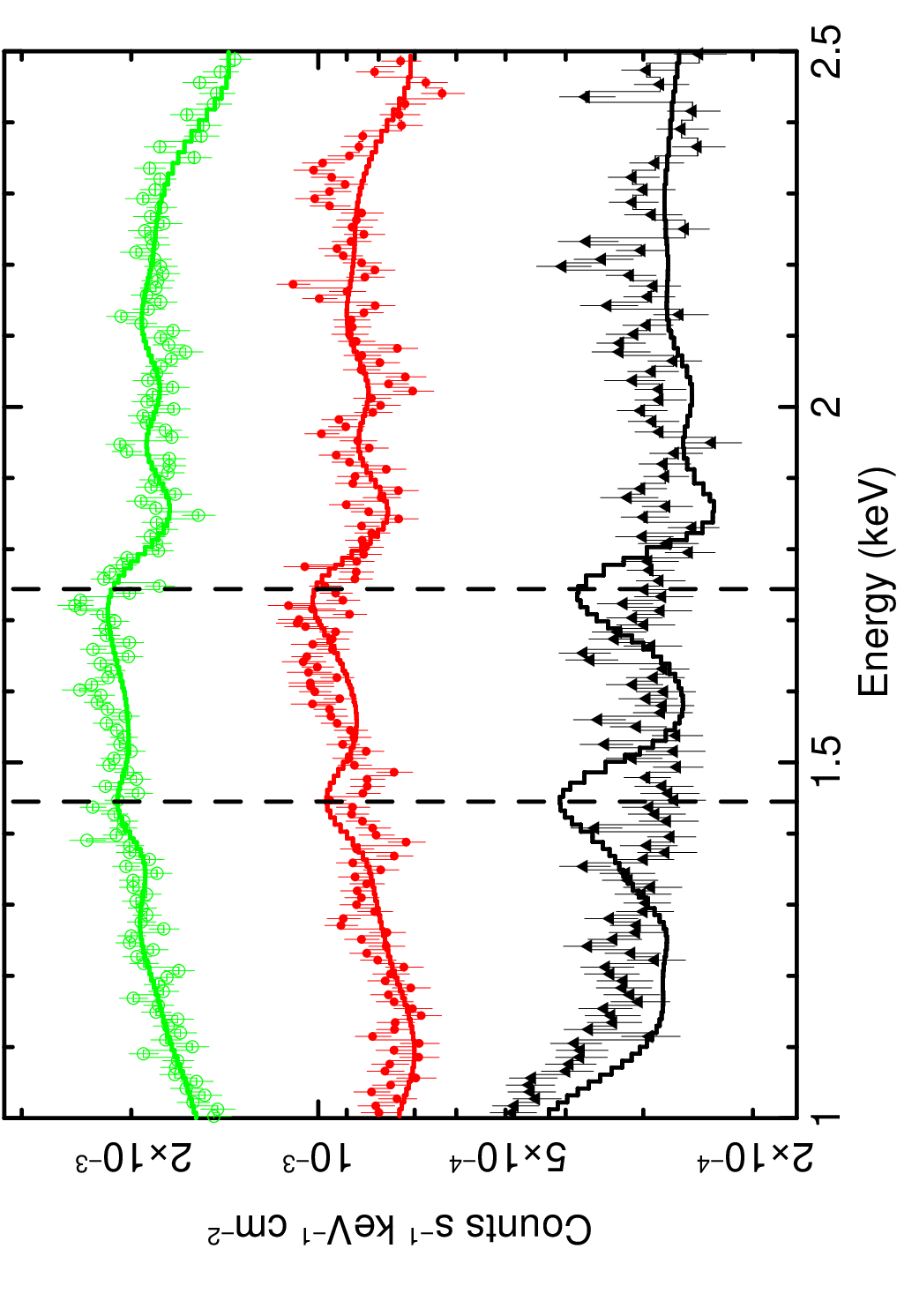}
}
 \figcaption[ev_pn]{Zoom into the Mg\,\textsc{xii}   energy range.  To highlight the variability of the emission lines (see  Table~\ref{tab:pnlines}) their intensities   were fixed to the best fit found for the last part of the observation.
Upper panel: RGS spectra of the first (filled circles and squares data points; blue and green in the electronic version) and last  intervals (open circles and squares data points; black and red in the electronic version).  Lower panel: EPIC-pn spectra for the three intervals (filled triangles interval A, filled squares interval B and open circles interval C).  The  vertical dashed lines indicate the Mg\,\textsc{xii}  Ly $\alpha$ ($E\sim1.45$ keV) and  the Mg\,\textsc{xii}  Ly $\beta $ ($E\sim1.74$ keV) emission lines, respectively.
 \label{fig:tab7}
}
\end{center}
\end{figure}

We note that  the Si and S   absorption features appear to be less prominent in the first two intervals, this is similar  to what  has been reported  for previous observations for  highly ionized iron (\citealt{Risaliti05,Brenneman13,Gofford2013,Maiolino10}):  when the  primary X-ray source is more absorbed by the  partial covering absorber the absorption lines are less prominent, possibly due to a lower continuum level.  We  found   the Mg\,\textsc{xii} Ly$\alpha$  and Ly$\beta$ emission lines to be weaker  at the beginning of the observation when the column density   of the partial coverer absorber is higher. The intensity of the   Mg\,\textsc{xii} Ly$\alpha$  varied  during the observation from   $I<0.7\times 10^{-6}$ ph cm$^{-2}$ s$^{-1}$ to  $I=2.1\pm 0.9 \times 10^{-5}$ ph cm$^{-2}$ s$^{-1}$   and a similar variability is also seen for the Mg \,\textsc{xii} Ly$\beta$ (see Table~\ref{tab:pnlines} and Figure~\ref{fig:tab7}). The variability of the  Mg\,\textsc{xii}  emission line is confirmed also by the inspection of the RGS data of the three intervals;  allowing its normalization to vary we found that, while in the third segment of the observation the Mg\,\textsc{xii}  Ly$\alpha$  emission line is clearly detected with a flux of $I=2.5\errUD{1.9}{1.4}\times 10^{-5}$ ph cm$^{-2}$ s$^{-1}$  at the beginning of the observation we can place only an upper limit  of $I< 0.7\times 10^{-5}$ ph cm$^{-2}$ s$^{-1}$. This latter value is in agreement with the intensity reported   from the analysis of the 0.5 Ms RGS data of \sorg\ ($I=0.5\pm0.2\times 10^{-5}$ ph cm$^{-2}$ s$^{-1}$; \citealt{Guainazzi09}). This is the opposite behaviour generally seen in the narrow emission lines that are associated with a distant emitter  (i.e. the narrow line region or the obscuring torus) whose intensity is consistent with being constant. Interestingly  in the RGS and EPIC-pn spectra of the third interval the Mg\,\textsc{xii} emission and absorption line profiles are like the classical P-Cygni profile, thus suggestive that these lines are produced in an outflowing absorber with a high covering factor, which we can associate with the mildly  ionized absorber (zone 2) . 
 
 \section{The location of warm absorbers}
During this X-ray monitoring campaign of \sorg\ we  witnessed  the  lowest $\nhsym$ and covering factor of the cold absorber  and  at the same time the source also increased  the intrinsic flux by a factor 10 with respect to the historical values; these two facts allowed us to unveil for the first time the presence of a low ionization outflowing wind in \sorg.  Here we briefly discuss the possible location of the   warm absorbers observed in \sorg. \\

 As already discussed in several works, \sorg\ is considered the best example of a variable partial covering absorber.  Prior to this monitoring program, the  column density of the cold X-ray absorber   was observed to vary from $\nhsym \sim10^{23}-10^{24}$ \nh\ (\citealt{Risaliti09b}). The  likely location of this variable cold absorber, as inferred from the time scale of the fast variability, was consistent with the BLR or the outer part of the BLR (at $R<10^{16}$ cm \citealt{Brenneman13,Maiolino10, Risaliti05}). 
Furthermore,  all the X-ray observations of \sorg\, except for  when the source is highly obscured,    showed  the presence of absorption lines due to a highly ionized wind (\citealt{Brenneman13,Risaliti05,Gofford2013}).  The detailed analysis of the spectral variability observed during the three deep Suzaku observations showed that the highly ionized component of the outflowing wind does not vary within the single observations but there is  significant variability on long time-scales (\citealt{Brenneman13}).  In particular  the ionization, the column  density   and the outflow velocity  varied in the two years gap  of the \suzaku\ observations between log $\xi\sim 3-4$ \logxi, $\nhsym \sim 10^{23}-10^{24}$ \nh\ and  $v\sim 2000-3600$ km s$^{-1}$. From the ionization equilibrium and the variability shown by this wind,  \citet{Brenneman13} inferred that the location of this wind is in the range of  $10^{14}-3\times 10^{15}$ cm   depending on the observation, which is in agreement with the measurements reported  from all the different X-ray observations of \sorg\ and within the location of the variable cold absorber. \\

 During the observation discussed here the highly ionized component of the absorber is characterized by a ionization state (log $\xi =3.8\pm 0.1$ \logxi), column density ($\nhsym =3.5\pm 0.9 \times  10^{23}$ \nh) and outflowing velocity ($v= 2200\errUD{300}{500}$ km s$^{-1}$),  which are in the range  reported in literature. 
 From these values  and using the relation between the intrinsic  continuum luminosity,  the ionization and the density of the absorber $\xi= L_{\mathrm{ion}}/ nR^2$  (\citealt{Tarter1969}) we can then place a limit on the location of this absorber. Assuming then an homogenous wind  and that  the thickness of the absorber is less than the distance from the central BH ($\Delta R/R < 1$) this  relation provides an upper limit on its distance   ($R< L_{\mathrm{ion}}/ \nhsym \xi$). As during  this observation the  intrinsic  luminosity was $ L(1-1000\; \mathrm{Rydberg}) \sim 10^{43}$ \lum\ (from extrapolating the best-fit   over the above energy range)  we derived a distance for this component  of $R<10^{16}$ cm; while the  outflowing wind  parameters are similar to the past measurements, the  higher value of the maximum distance of this wind is driven by the higher luminosity  with respect to the past observations.  As the  estimates for the mass of the super massive black hole  in \sorg\ range from $M_{\mathrm{BH}}\sim 2\times 10^6$ M$_\odot$ to $10^{8}$ M$_\odot$ (\citealt{Risaliti09b}), the inferred maximum  distance  correspond to $300-16000\;  r_{g}$, placing the wind   within the outer part of the BLR.  The  minimum launch radius of this wind can be derived assuming that  the outflow velocity  across the line of sight is equal to the escape velocity at the observed radius; from the best fit model  we found $R_\mathrm{min}>10^{4}$ $r_\mathrm{g}$.  As there is rather wide range for the  black hole mass of \sorg\  in the following we adopt  a mean value of $ 10^{7}$   M$_\odot$.   \\
 
  We  confirm  the well known  variability of  the partial covering low ionisation absorber; in particular we measured a  variability of the column density of about $\Delta N_\mathrm{H}\sim10^{23}$ \nh\ during the 130 ks exposure.  Thus  assuming that the  obscuring clouds are moving with Keplerian velocity ($v_\mathrm{k}$) a limit on the size of these clouds $\Delta R$ can be obtained  using the relation $\Delta R=v_{\mathrm k} \Delta t$ and the duration of the observation ($\Delta t\sim 130$ ks) for  the elapsed  time  for the passage of the clouds.   If  the partial covering  absorber is the same responsible for the $\feka$ emission line, we can then  obtain an estimate of the clouds velocity from the  $FWHM$ of the $\feka$.   As during the \chandra\  observation \sorg\ was in a Compton-thick state,  the high resolution  grating data provide the best measurement of width of the $\feka$ narrow core, which is  $\sigma= 30\errUD{24}{11}$ eV (Gofford et al. in prep.) corresponding to a $FWHM \sim 3000$ km s$^{-1}$, in agreement with the value measured with the past \xmm\ observations (\citealt{Risaliti2002}).  
 Assuming a Keplerian velocity,  this   would correspond to a distance of these clouds   of  about   a $10^4$ $r_{g}$ and thus of the same order of the distance of the outer BLR and with a cloud size of $\Delta R <  10^{14}$ cm.  Finally, a first order estimate of the density of these clouds can be derived from  the relation     $n=\nhsym/\Delta R \sim10^{9}$ cm$^{-3}$, which is also in the typical range of the BLR clouds density. 
Interestingly the RGS data  suggest that this variable absorber is a low ionization component of the ionized outflowing wind characterized by log$\xi=0.63\pm0.03$ \logxi\ and a $\nhsym\sim 7\times 10^{22}$ \nh\ (see Table~4 and Table~6).  From  the estimated  size of these clouds ($\Delta R \sim 3\times 10^{13}$ cm),  the  density of this low ionized component of the wind  and recalling that       $R^2= L_{\mathrm{ion}}/ n\xi$   we  can derive a first order estimate of the   location of this variable ionized absorber of $R<10^{17}$ cm. This estimate would then place again this low ionization absorber at  a distance of the same order of the outer BLR and  coincident with the location of the highly ionized outflowing wind.  A possible scenario is that  this zone  is just   the outer part of the wind and when its column density increases and or ionization decreases it could be responsible for covering-up  the strong broad Mg \,\textsc{xii} emission lines   and the X-ray source. The uncovering of \sorg, that we witnessed at the end of this observation, might be explained if we are seeing through a hole or a gap in this patchy lowly ionized absorber which is characterized  by   $\nhsym = 0.8\pm 0.2 \times 10^{22}$ \nh.  
 
 The most interesting and new result of this  \xmm\ observation of \sorg\ has been the discovery  of a mildly  (log $\xi=2.1\pm 0.1$ \logxi\  see Table~4) ionized component of the outflowing wind, which is responsible  for the Mg \,\textsc{xi}  and Mg \,\textsc{xii}  absorption features. In order to place more constraints on the location  of this ionized absorber, we  investigated any possible   variability of its  column density or ionization during the observation. However, any column density  or ionization variability  is strongly degenerate with the variability of the partial covering absorber.   From the ionization  parameter of this mildly ionization component of the wind (zone 2),  its column density  and using again   the relation  $R< L_{\mathrm{ion}}/ \nhsym \xi$  we can derive only a relatively loose constraint on its distance  as $R_\mathrm{max}< 10^{19}$ cm  (see Table~\ref{tab:ionabs}). This estimate  corresponds to a location outside the BLR and more typical of the putative obscuring torus; but as we discuss below  the most plausible location of this absorber is   closer in,  at  a distance smaller than the location  of the  variable partial covering absorber  and at a similar distance of the highly ionized component    (i. e. $R < 10^{16}$ cm). \\ 
  
 As we are looking towards the innermost region  through  a partial covering ionized absorber (zone 3), we might  be able to see the  mildly ionized  disk wind that imprints its features on the primary continuum,  but  only when the covering factor  and  $\nhsym$   of the partial coverer are sufficiently low. 	Following this argument the most likely location of this mildly ionized absorber would  then be within the   outer part of the BLR. 
This scenario would naturally explain why  this wind, like the highly ionized outflow, is not seen when the column density of the low ionization  partial covering  absorber increases, as then this part of the AGN  is shielded from view.   This scenario is also in agreement  with the location derived from its  velocity:  from  $v_\mathrm{out}=1700\pm500$ km s$^{-1}$ we derive  $R_\mathrm{min}> 3\times 10^{4}$ $r_\mathrm{g}$ which for a black hole mass of $ 10^{7}$   M$_\odot$  corresponds to   $R_\mathrm{min}>10^{16}$ cm.  \\

Another argument that points towards a location within the partial coverer  is the detection,  when the source is less obscured, of    possible P-Cygni profiles for the Mg\,\textsc{xi} and Mg\,\textsc{xii}  lines, which can be interpreted as arising  from the zone 2 outflowing wind.  As discussed in \S ~3.3 the similarity  of the  strengths  of the emission and absorption components of the P-Cygni profile indicates a covering factor close to unity  of this wind, while the   widths of these  P-Cygni lines  ($\sigma= 9 \pm 3$ eV or $FWHM \sim4300 $ km s$^{-1}$), if interpreted  as velocity broadening would naturally place  this wind within the  putative torus  and thus closer in than the classical warm absorbers ($R\sim  5\times 10^{3}$ $ r_\mathrm{g}$ or $R\sim10^{16}$ cm  for a black hole mass of  $M_{\mathrm{BH}}\sim   10^7$ M$_\odot$).
 Furthermore, the inspection of the three  RGS  and EPIC-pn spectra provides a tentative evidence of variability, during  the single \xmm\ observation (see Table~7), of the  intensity of these  Mg\,\textsc{xii}  emission  lines. As the  emission line emerges when the source is brighter and less obscured we are  facing two possible scenarios: if we place this gas  within the partial covering  absorber (zone 3) this could be simply explained as an effect of the uncovering of the primary  AGN emission. On the other hand  if we place the zone 2 gas   outside the partial covering absorber then the increase of the intensity  of these emission lines  could be explained  due to a higher illuminating continuum  leaking through the  absorber. However, assuming that the line is responding to the  illuminating  continuum, the short time scale of this response implies a compact region that again would  place this ionized gas within the variable X-ray absorber (zone 3); as the line is responding within the duration of the observation ($\Delta t \sim 130$ ks)  a simple estimate  for the distance of the emitting gas based on the light crossing time gives a $R\sim 10^{15}$ cm.  Thus we conclude that  the detection of this feature most likely implies a sub parsec distance for this ionized gas.\\

We can now estimate the mass outflow rate and energetics of this wind and compare them with the properties of the higher ionization wind. The mass outflow rate can be estimated using the expression $\dot M_{\mathrm{out}} = \Omega \, m_\mathrm{p}\, \nhsym\, v_{\mathrm{out}} \,R$ (\citealt{Krongold2007}), which is appropriate for a  biconical wind, where $\Omega$ is the  covering factor of the wind (in terms of solid angle). For the most conservative  estimate of the wind energetics we adopt the outflowing velocity as measured with the \textsc{xstar} modeling  of the RGS data ($v_{\mathrm{out}}= 1700\pm 500$ km s$^{-1}$).  For the distance we recall that from the ionization parameter and column density  we derived an upper limit on the distance of $R_{\mathrm{max}}< \mathrm {few} \times 10^{19}$ cm;  while  from the velocity  broadening  of the absorption lines and assuming the mean mass  we can put  a lower limit  for this radius of about $R_{\mathrm{min}} \sim 5 \times 10^3$ $r_\mathrm{g}$. Given that the Mg\,\textsc{xii} emission and absorption  lines are reminiscent of a P-Cygni profile, it is plausible  that the mildly ionized   (log $\xi \sim 2$ \logxi) component of the wind  is also responsible for the  Mg\,\textsc{xii}  Ly$\alpha$ emission line. Thus we  estimated the global  covering fraction, as a fraction of $4\pi$ sr, for this absorber to produce the total observed luminosity  of    the Mg\,\textsc{xii} Ly$\alpha$ emission line of $L\sim 3\times 10^{39}$ erg s$^{-1}$  (from $I_{\mathrm{em}}\sim 2\times 10^{-5}$ photons cm$^{-2}$ s$^{-1}$; see Table~\ref{tab:pnlines}).
We used the \textsc{xstar} code to calculate the Mg\,\textsc{xii} line luminosity for a spherical shell of gas, covering  a full $4\pi$ sr and  illuminated by the ionizing luminosity of \sorg\ ($L_{\mathrm{1-1000\; Rydbergs}}\sim 1.5\times 10^{43}$ \lum), which  was derived  extrapolating the best fit over the $1-1000\; \mathrm {Rydbergs}$ energy range.   Adopting the  column density and ionization parameter  as measured for  the low ionization absorber ($\nhsym \sim   10^{22}$  \nh\ and log $\xi\sim 2$ \logxi), we found that this absorber would produce a Mg\,\textsc{xii} luminosity  of $ 6.3 \times 10^{39}$\lum\ for a  full $4\pi$ sr   shell of gas covering the AGN. Thus, in order to reproduce the   measured luminosity  of  $\sim 3 \times 10^{39}$ \lum,   this absorber would require a covering fraction of about $2 \pi $. \\

Thus for a column density of the wind of $\nhsym\sim 1\times 10^{22}$ \nh\ and the above estimate of the covering fraction and distance   of  $R\sim 10 ^{16}$ cm (from  equating the  wind velocity to the escape velocity), we conservatively estimate  $\dot M_{\mathrm{out}}\sim 2 \times 10^{23}$ g s$^{-1}\sim 3\times 10^{-3}$ M$_\odot$ yr$^{-1}$ or $\dot M_{\mathrm{out}}/\dot M_{\mathrm{Edd}}\sim 0.01$ (assuming  the accretion efficiency $\eta=0.06$), which  means that the outflow rate is only few percent of the Eddington   mass accretion rate.  Assuming that the wind has reached the terminal velocity, we can also calculate the mechanical power of the wind $\dot E=1/2\dot M_{\mathrm{out}}\, v^2_{\mathrm{out}}\sim  10^{40}$ \lum\  and compare it  with the $L_{\mathrm{Edd}}\sim 1.3\times 10^{45}$ \lum,   and the bolometric luminosity of \sorg\ ($L_{\mathrm{Bol}}\sim 10^{44}$ \lum\ ,  \citealt{Vasudevan2010}). The kinetic power of the wind is only $\sim 0.01$\%  of the bolometric luminosity. 
For the highly ionized  wind following the same arguments and assuming  that it has the same covering factor and  the averaged wind velocity of $v_{\mathrm{out}}\sim  2300$ km s$^{-1}$ (see Table~\ref{tab:ionabs_pn}), we estimate a mass outflow rate of $\dot M_{\mathrm{out}}\sim  8\times 10^{-2}$ M$_\odot$ yr$^{-1}$ and a mechanical power of $\dot E=1/2\dot M_{\mathrm{out}}\, v^2_{\mathrm{out}}\sim  10^{41}$ \lum, which albeit the large uncertainties is of about  $\sim 0.1$\%  of the bolometric luminosity.  
 In both cases the wind energetics are thus below  the typical values necessary for a significant feedback even considering the lower value obtained  in the most  recent simulation  (0.5--5\%  \citealt{HopkinsElvis2010,DiMatteo2005,Tombesi2013}). Nevertheless,   the mass loss rates $\dot M_{\mathrm{out}}\sim  0.3-8\times 10^{-2}$ M$_\odot$ yr$^{-1}$ of the   medium and highly ionized components   are comparable or slightly above  the mass accretion rate  of \sorg\  that  can be  derived from the bolometric luminosity  ($\dot M_{\mathrm{acc}}\sim 2 \times 10^{24}$ g s$^{-1}\sim 3\times 10^{-2}$ M$_\odot$ yr$^{-1}$,  assuming  $\eta=0.06$).  Although these estimates  strongly depend on the covering factors derived from the Mg\,\textsc{xii} emission  line   and the \,\textsc{xstar} component,  they  suggest that  both these ionized winds play an important role in the  energy and mass budgets of the active nucleus and  can nevertheless  have a feedback for the innermost region of the host galaxy.  \\

\section{Conclusions}
We  presented the results of a detailed analysis of one   \xmm\  observation of \sorg\,  which caught \sorg\ in a high flux state characterized also by the lowest covering factor ($\sim 30$\%) and column density  of the partial covering X-ray absorber ever recorded.  Thanks to the combination of the higher flux and lower $\nhsym$ we could perform a time resolved spectral analysis  of the \xmm-RGS data, which unveiled a variability  of this low column density absorber on a time-scale as short as 40 ks. One of the main results of this  observation is   that the  ionized  disk wind present in \sorg\  is complex and  composed by at least two ionization zones,  possibly three as suggested by the presence of several additional absorption lines both in the RGS and the EPIC-pn spectra.  
Besides the well studied highly ionized component, which is responsible for the He- and H-like  Fe absorption features seen in the EPIC-pn X-ray spectra of \sorg,   we detected a second zone with log $\xi \sim 2$ \logxi\ responsible for the Ne, Mg and Si absorption lines.  The presence of an additional absorption feature at $E\sim 0.76$ keV suggests a  third lower ionization zone with log $\xi < 1$ \logxi and   we propose that the variable partial covering absorber located at the BLR distance  could be  identified with  this low ionization zone of the disk wind.
We also detected two possible P-Cygni profiles, where the absorption components  have  widths and strengths  similar to the  corresponding emission components, suggestive of a wide angle outflowing wind.  From the broadening of the profile and   variability of the emission lines we  suggest that  this wind is located within the variable soft X-ray absorber.\\

  We would like to thank the anonymous referee for his/her useful comments, which have improved this paper. This paper has made use of observations obtained with \xmm\, an ESA science mission with instruments and contributions directly funded by ESA member states and the USA (NASA) and with the  \chandra\  X-ray Observatory.  This research has made use of software provided by the Chandra X-ray Center (CXC) in the application packages CIAO, ChIPS. J.N. Reeves acknowledges Chandra grant  GO2-13123A. G. Risaliti  acknowledges the NASA grant NNX13AH71G and  financial support from  the Italian grant PRIN-INAF 2012.

\newpage 

\newpage

\begin{deluxetable}{ccccc}
\tablecaption{Log of the observations and exposure times.}
\tablewidth{0pt}
\tablehead{
\colhead{Mission} & \colhead{Instrument} & \colhead{T$_{\rm(total)}$ (ks)} & \colhead{T$_{\rm(net)}$ (ks)}
& \colhead{DATE} 
}
  \startdata  
 \xmm\ & PN& 132.2 &92.7	&2013-01-23 \\
  \xmm\ & RGS1&  	133.7 &  113.2	&2013-01-23 \\
 \xmm\ & RGS2& 133.7 &113.3	&2013-01-23 \\
 \chandra\ & ACIS-S HETG& 90&-&	2012-04-09 \\
 \chandra\ & ACIS-S HETG& 110&-&	 2012-04-12 \\
\enddata
\tablecomments{For \xmm\ the net exposure times are  after  filtering for high-background time intervals.}
\label{tab:log_observ}
\end{deluxetable}

\begin{deluxetable}{cccccc}
 \tablecaption{\xmm\ RGS1 and RGS2: best fit absorption lines  required in the third interval. The  statistic   for the model with no lines is $C/d.o.f. = 1290.6/971$}
\tablewidth{0pt}

\tablehead{
\colhead{Rest Energy$^{\mathrm{a}}$}     & \colhead{Intensity} & \colhead{$\sigma^{\mathrm{b}}$} & \colhead{ID$^{\mathrm{b}}$}      & \colhead{Atomic Energy$^{\mathrm{d}}$}& \colhead{$\Delta C^{\mathrm{e}}$}\\
   \colhead{(eV)} & \colhead{($10^{-5}$ph cm$^{-2}$ s$^{-1}$)}&  \colhead{(eV)} & & \colhead{(eV)} & \\
  }
   \startdata    
760.4$\errUD{5.7}{5.4}$&$ -1.57\errUD{0.76}{0.70}$ &$7\pm 5$ & Fe\,\textsc{xvii} -\textsc{xviii}& ... & 10.8\\   
925.7$\errUD{2.8}{2.3}$& $-1.39\errUD{0.59}{0.79}$ &$<4$ & Ne\,\textsc{ix}& 922.0 & 11.4\\   
971.6$\errUD{1.6}{1.9}$& $-1.95\errUD{0.73}{0.46}$ &$<4$ & Fe\,\textsc{xx}$2p-3d$& 967.3 & 16.9\\   
1355.4$\errUD{7.9}{7.6}$&$-3.46\errUD{1.26}{1.21}$ &$9\pm3$ & Mg\,\textsc{xi}& 1352.2 & 13.5\\   
1476.5$\errUD{3.9}{4.9}$&$-7.46\errUD{1.55}{1.47}$ &$9^{\mathrm{f}}$ & Mg\,\textsc{xii}& 1472.6 & 17.5\\   
  \enddata
\tablecomments{The $\Gamma$ of the soft power-law is tied to the  hard power-law component and is fixed to the best fit value for the EPIC-pn spectrum.\\ ${^\mathrm{a}}$ Measured line energy in the  \sorg\ rest frame.\\ $^{\mathrm{b}} 1\sigma$ width of the absorption line in eV. \\$^{\mathrm{c}}$ Possible identification. \\$^{\mathrm{d}}$ Known atomic energy of the most likely identification of the  line in eV.   \\
$^{\mathrm{e}}$ Improvement in C-statistic upon adding the line.\\$^{\mathrm{f}}$ Tied to the Mg\,\textsc{xi} line width.\\}
 \label{tab:absline}
\end{deluxetable}

\begin{deluxetable}{cccccc}
\tablecaption{\xmm\ RGS1 and RGS2: best fit  parameters of the strongest emission lines  required in the third interval.}
\tablewidth{0pt}

\tablehead{
\colhead{Rest Energy$^{\mathrm{a}}$}     & \colhead{Intensity}  & \colhead{ID$^{\mathrm{b}}$}      & \colhead{Atomic Energy$^{\mathrm{c}}$} &\colhead{$\Delta C^{\mathrm{e}}$}\\
   \colhead{(eV)} & \colhead{($10^{-5}$ph cm$^{-2}$ s$^{-1}$)}&   & \colhead{(eV)} &  \\
  }
   \startdata    
499.9$\errUD{0.5}{0.5}$&$ 2.56\errUD{0.68}{1.38}$  & N\,\textsc{vii}& 500.4& 16.9\\   
560.5$\errUD{0.8}{0.8}$& $3.12\errUD{1.43}{1.39}$   & O\,\textsc{vii}&  561.0&13.6\\   
653.2$\errUD{1.7}{1.5}$& $0.98\errUD{0.76}{0.69}$  & O\,\textsc{viii}&  653.7 &9.2\\   
 1462.0$\errUD{4.9}{5.8}$&$6.54\errUD{1.61}{1.51}$ &  Mg\,\textsc{xii}& 1472.6&11.3  \\   
  \enddata
\tablecomments{${^\mathrm{a}}$ Measured line energy in the  \sorg\ rest frame.\\$^{\mathrm{b}}$ Possible identification. \\$^{\mathrm{c}}$ Known atomic energy of the most likely identification of the  line in eV.   \\}
 \label{tab:emlinergs}
\end{deluxetable}

\begin{deluxetable}{cccc}
\tablecaption{\xmm\ RGS and EPIC-pn best fit parameters for the ionized absorber model. The  statistics  for  the model are $C/d.o.f.=1175.0/965$ and $\chi^2/d.o.f.=1292.2/1253$ for  the RGS and the EPIC-pn spectra respectively.}
\tablewidth{0pt}
\tablehead{
\colhead{Component}      &\colhead{Parameter}      & \colhead{RGS} & \colhead{EPIC-pn}  \\
  }
   \startdata    

Power-law & $\Gamma$ & $2.1^{\mathrm{f}} $&2.05\errUD{0.04}{0.06} \\

&$A_{\mathrm{scattered}}^\mathrm{{a}}$& $2.4 \errUD{0.8}{0.9}$ &3.1\errUD{1.3}{1.6} \\

& $A_{\mathrm{primary}}^\mathrm{{b}}$&$2.1 \errUD{0.2}{0.2}$&1.5\errUD{0.1}{0.2} \\

ionised absorber &$\nhsym^\mathrm{{c}}$&-&21.5\errUD{11.8}{7.8} \\
zone 1& log $\xi^\mathrm{{d}}$&-&3.77 \errUD{0.12}{0.09} \\
& $v_{\mathrm{out}}^\mathrm{{e}}$&-&3900\errUD{800}{1000} \\

ionised absorber &$\nhsym^\mathrm{{c}}$&$1.1\errUD{0.4}{0.3}$&1.0\errUD{0.3}{0.2} \\
zone 2 & log $\xi^\mathrm{{d}}$&$2.15\errUD{0.12}{0.10}$&2.11\errUD{0.16}{0.08} \\

& $v_{\mathrm{out}}^\mathrm{{e}}$ & $1700\errUD{400}{500}$ &= RGS\\
Fully covering absorber & $\nhsym^\mathrm{{c}}$ & $<0.1$ &0.3\errUD{0.3}{0.1} \\
&&&\\
Partial covering &&& \\
ionised absorber &$\nhsym^\mathrm{{c}}$&1.1\errUD{0.1}{0.1}&1.3\errUD{0.2}{0.3} \\
zone 3& log $\xi^\mathrm{{d}}$&0.17\errUD{0.08}{0.09} &0.6 \errUD{0.2}{0.3} \\
&CF  &  1$^{g}$& $0.94\pm0.04$   \\

$$
\enddata

\tablecomments{\\
$^{\mathrm{a}}$ Normalization in units of $10^{-4}$ photons cm$^{-2}$ s$^{-1}$.\\
$^{\mathrm{b}}$ Normalization in units of $10^{-2}$ photons cm$^{-2}$ s$^{-1}$.\\
$^{\mathrm{c}}$ $\nhsym$ in units of $10^{22}$ \nh.\\
$^{\mathrm{d}}$ Log of the ionization parameter,  in units of erg cm s$^{-1}$.  \\
$^{\mathrm{e}}$ Outflow velocity in km s $^{-1}$.\\
$^{\mathrm{f}}$ Denotes the parameter is fixed.\\
$^{\mathrm{g}}$ As the RGS has a limited bandpass we assumed a covering fraction of unity for the zone 3 absorber.\\
}
 \label{tab:ionabs}
\end{deluxetable}

\begin{deluxetable}{ccccc}
\tablecaption{\xmm\  EPIC-pn soft X-ray absorption  lines. }

\tablehead{
\colhead{Rest Energy$^{\mathrm{a}}$}     & \colhead{Intensity} & \colhead{ID$^{\mathrm{b}}$}      & \colhead{Atomic Energy$^{\mathrm{c}}$}& \colhead{$(\Delta \chi^\mathrm{2})^\mathrm{d}$}\\
   \colhead{(keV)} & \colhead{($10^{-5}$ph cm$^{-2}$ s$^{-1}$)}&  & \colhead{(keV)} & \\
  }
   \startdata    
$0.77\errUD{0.03}{0.03}$&$-3.5\errUD{1.6}{1.5}$& Fe\,\textsc{xvii} -\textsc{xviii} &...&14.2\\
$1.33\errUD{0.02}{0.03}$&$-3.0\errUD{0.9}{0.9}$& Mg\,\textsc{xi}  &1.33&18.0\\
$1.47\errUD{0.02}{0.02}$&$-3.1\errUD{0.9}{1.0}$& Mg\,\textsc{xii} &1.47&16.9\\
$1.84\errUD{0.02}{0.02} $& $-4.0\errUD{1.1}{1.0}$  & Si\,\textsc{xiii}& 1.84  & 37.4\\   
$2.49\errUD{0.04}{0.03}$&$-2.2\errUD{0.9}{1.0}$  & S\,\textsc{xv} &2.46 & 16.7\\   
$6.77\errUD{0.03}{0.02}$&$-2.2\errUD{0.4}{0.4}$& Fe\,\textsc{xxv} &6.63&66.4\\
$7.05\errUD{0.03}{0.02}$&$-2.1\errUD{0.5}{0.4}$& Fe\,\textsc{xxvi} &6.97&60.7\\

  \enddata
\tablecomments{ ${^\mathrm{a}}$ Measured line energy in the  \sorg\ rest frame.\\ $^{\mathrm{b}}$ Possible identification. \\$^{\mathrm{c}}$ Known atomic energy of the most likely identification of the  line in eV.    \\${^\mathrm{d}}$ {$(\Delta \chi^\mathrm{2})$  with respect to a fit without the two ionized absorbers, which provides  $\chi^2/d.o.f=1500.7/1258$}\\}
  \label{tab:pnabslines}
\end{deluxetable}

\begin{deluxetable}{ccccc}
\tablecaption{\xmm\ -EPIC-pn best fit parameters for the 3  ionized absorbers model. The  statistics  for  the model is $\chi^2/d.o.f.=3909.3/3664$.}
\tablewidth{0pt}
\tablehead{
\colhead{Component}      &\colhead{Parameter}      & \colhead{A} & \colhead{B} & \colhead{C} \\
  }
   \startdata    

Power-law  			&$A_{\mathrm{Scattered}}^\mathrm{{a}}$	& 	$2.1 \errUD{0.1}{0.1}$ 	&	3.1\errUD{0.1}{0.1} &	 5.1\errUD{0.2}{0.2}  \\
		   			& $A_{\mathrm{primary}}^\mathrm{{b}}$	     &	$0.62 \errUD{0.04}{0.04}$  &0.84\errUD{0.04}{0.02}   & 1.22\errUD{0.05}{0.06} \\
Fully covering absorber & $\nhsym^\mathrm{{c}}$ &2.0\errUD{0.2}{0.1} &0.96\errUD{0.06}{0.07} & 0.84\errUD{0.09}{0.05}\\

&&&\\

ionised absorber 	&$\nhsym^\mathrm{{c}}$						&35.3\errUD{8.1}{8.9}  		&35.3$^{t}$				& 35.3$^t$ \\
zone 1				& log $\xi^\mathrm{{d}}$						& 3.81\errUD{0.05}{0.06} 	& 3.81$^t$ 				&3.81$^t$ \\
& $v_{\mathrm{out}}^\mathrm{{e}}$									 & 2300\errUD{500}{400}		& 2300$^{t}$ 		& 2300$^t$\\
&&&\\

ionised absorber 	&$\nhsym^\mathrm{{c}}$ 						&1.7\errUD{0.3}{0.3} 		& 1.7$^t$			&  1.7 \\
zone 2 				& log $\xi^\mathrm{{d}}$						&2.3\errUD{0.1}{0.1} 	& 2.3$^{t}$ 				&2.3$^{t}$ \\

&&&\\
Partial covering &&& \\
ionised absorber &$\nhsym^\mathrm{{c}}$&7.0\errUD{0.4}{0.5}&2.4\errUD{0.1}{0.1} &0.8\errUD{0.1}{0.2} \\
zone 3& log $\xi^\mathrm{{d}}$&0.63\errUD{0.03}{0.02} &0.63$^{t}$ 				&  0.63$^{t}$ \\
&CF  &$0.87\pm0.01$ & 0.87$^{t}$ &0.87$^{t}$ \\
\enddata

\tablecomments{\\
$^{\mathrm{a}}$ Normalization, in units of $10^{-4}$ photons cm$^{-2}$ s$^{-1}$.\\
$^{\mathrm{b}}$ Normalization, in units of $10^{-2}$ photons cm$^{-2}$ s$^{-1}$.\\
$^{\mathrm{c}}$ $\nhsym$ in units of $10^{22}$ \nh.\\
$^{\mathrm{d}}$ Log of the ionization parameter,  in units of erg cm s$^{-1}$.  \\
$^{\mathrm{e}}$ Outflow velocity in km s $^{-1}$.\\
 $^{\mathrm{g}}$ Fully covering absorber.\\
$^{\mathrm{t}}$ Denotes the parameter is tied.}
 \label{tab:ionabs_pn}
\end{deluxetable}

\begin{deluxetable}{lccc}
\tablecaption{Intensities of the Mg \,\textsc{xii}  emission lines}  \tablehead{
\colhead{} & RGS & \multicolumn{2}{c}{EPIC-pn} \\
\hline
\colhead{Interval}&\colhead{Mg\,\textsc{xii} Ly$\alpha$} &Mg\,\textsc{xii} Ly$\alpha$   &\colhead{Mg\,\textsc{xii} Ly$\beta$}\\
   & \colhead{($10^{-5}$ph cm$^{-2}$ s$^{-1}$)}&    \colhead{($10^{-5}$ph cm$^{-2}$ s$^{-1}$)} & \colhead{($10^{-5}$ph cm$^{-2}$ s$^{-1}$)}\\
  }
   \startdata    
 INT A& $<0.7$& $<0.07$ & $<0.3$\\
 INT B&$<1.3$ & $<0.3$&$<0.9$\\
 INT C&$2.5\errUD{1.9}{1.4}$ &$2.1\errUD{0.9}{0.9}$&$1.9\errUD{1.6}{1.3}$\\

  \enddata
  \label{tab:pnlines}
\end{deluxetable}
\end{document}